\documentclass{ieeeaccess}

\usepackage[T1]{fontenc}
\usepackage{times}
\usepackage{helvet}

\usepackage[rgb]{xcolor}
\usepackage{amsmath,amssymb,amsfonts}
\usepackage{algorithmic}
\usepackage{graphicx}
\usepackage{textcomp}
\usepackage{cite}
\usepackage{multirow}
\usepackage{tabularx}
\usepackage{array}
\usepackage{hyperref}
\usepackage{longtable}
\usepackage{pdflscape}
\usepackage{float}
\usepackage{placeins}
\usepackage{cuted}
\usepackage{bm}

\def\BibTeX{{\rm B\kern-.05em{\sc i\kern-.025em b}\kern-.08em
    T\kern-.1667em\lower.7ex\hbox{E}\kern-.125emX}}

\begin{document}
\history{Received 1 December 2025, accepted 17 December 2025, date of publication 22 December 2025, date of current version 2 January 2026.}
\doi{10.1109/ACCESS.2025.3646769}

\title{Architectural Selection Framework for Synthetic Network Traffic: Quantifying the Fidelity-Utility Trade-off}
\author{\uppercase{Dure Adan Ammara}\authorrefmark{1}, 
\uppercase{Jianguo Ding}\authorrefmark{1}\IEEEmembership{Senior Member, IEEE} , and Kurt Tutschku\authorrefmark{1}
\IEEEmembership{Member, IEEE}
}

\address[1]{Blekinge Institute of Technology, Karlskrona, 37179 Sweden}

\tfootnote{This work was supported by European Celtic+ and Swedish Vinnova Project through CISSAN–Collective Intelligence Supported by Security Aware Nodes under Grant C2022/1-3.}

\markboth
{Author \headeretal: Preparation of Papers for IEEE TRANSACTIONS and JOURNALS}
{Author \headeretal: Preparation of Papers for IEEE TRANSACTIONS and JOURNALS}

\corresp{Corresponding author: Dure Adan Ammara (dure.adan.ammara@bth.se).}

\begin{abstract}
The fidelity and utility of synthetic network traffic are critically compromised by architectural mismatch across heterogeneous network datasets and prevalent scalability failure. This study addresses this challenge by establishing an Architectural Selection Framework that empirically quantifies how data structure compatibility dictates the optimal fidelity-utility trade-off. We systematically evaluate twelve generative architectures (both non-AI and AI) across two distinct data structure types: categorical-heavy NSL-KDD and continuous-flow-heavy CIC-IDS2017. Fidelity is rigorously assessed through three structural metrics (Data Structure, Correlation, and Probability Distribution Difference) to confirm structural realism before evaluating downstream utility. Our results, confirmed over twenty independent runs ($N=20$), demonstrate that GAN-based models (CTGAN, CopulaGAN) exhibit superior architectural robustness, consistently achieving the optimal balance of statistical fidelity and practical utility. Conversely, the framework exposes critical failure modes, i.e., statistical methods compromise structural fidelity for utility (Compromised fidelity), and modern iterative architectures, such as Diffusion Models, face prohibitive computational barriers, rendering them impractical for large-scale security deployment. This contribution provides security practitioners with an evidence-based guide for mitigating architectural failures, thereby setting a benchmark for reliable and scalable synthetic data deployment in adaptive security solutions.

\end{abstract}

\begin{keywords}
Synthetic Data Generation, Generative Adversarial Networks (GANs), NSL-KDD, CIC-IDS, Network Traffic Analysis, Fidelity, Utility, Generative AI
\end{keywords}

\titlepgskip=-21pt

\maketitle

\section{Introduction}
\label{sec1}

The increasing reliance on data-driven decision-making in networking and cybersecurity has intensified the need for high-quality network traffic data. This data plays a fundamental role in various applications, including intrusion detection, threat analysis, and the training of artificial intelligence (AI) models for adaptive security solutions \cite{shahraki2022comparative}. However, collecting and utilizing real-world network traffic data remains a significant challenge due to privacy restrictions, limited availability, and the high cost of manual labeling \cite{guerra2022datasets}. These barriers hinder the development and evaluation of robust AI-based security systems.

\subsection{Need for High-Fidelity Network Data}

Synthetic data generation has emerged as a promising solution to overcome the challenges mentioned above. It enables the creation of realistic yet privacy-preserving network data that can supplement or replace real traffic for research and testing. The primary challenge synthetic data must address is the imbalance issue: real-world traffic datasets are highly skewed, with benign flows vastly outnumbering malicious ones. This imbalance biases AI models against rare but critical attack classes \cite{layeghy2024benchmarking}. Moreover, regulatory constraints (e.g., GDPR) and organizational confidentiality further limit access to comprehensive network traces \cite{kotal2022privetab}. Hence, generating high-fidelity and balanced synthetic traffic is essential for scalable, ethical, and reproducible cybersecurity research \cite{ganji2023towards}. Despite this potential, generating high-quality synthetic network traffic data remains a technically complex task. Existing methods struggle with three core issues:

\begin{enumerate}
    \item Ensuring Realism (Fidelity): replicating the complex statistical and temporal dependencies of real network traffic \cite{sarker2020cybersecurity}.
    \item Variability and Generalization (Utility): avoiding mode collapse to produce diverse traffic variations effective for AI training \cite{arjovsky2017wasserstein,rao2023imbalanced,lin2022idsgan}.
    \item Computational Complexity: managing the resource demands of state-of-the-art models like Generative Adversarial Networks (GANs) and diffusion models \cite{sivaroopan2023synig}.
\end{enumerate}
These challenges are not independent; they are fundamentally influenced by the underlying architecture of the generative model. In particular, a model’s ability to maintain fidelity, achieve utility, and manage computational complexity depends on how well its architecture aligns with the data’s intrinsic structure.
\subsection{Motivation}
Despite significant advances in generative modeling, the motivation for this study stems from a critical gap between model innovation and model applicability: the lack of architectural understanding that connects generative design to data structure. While these challenges are well recognized, most existing studies focus on improving individual generative models rather than understanding \textit{why} specific architectures succeed or fail across different network data structures. Network traffic data is inherently heterogeneous; some datasets (e.g., NSL-KDD) are dominated by categorical features, while others (e.g., CIC-IDS2017) are continuous and high-dimensional. Without systematic guidance on aligning model architectures to these varying data characteristics, synthetic data generation remains inconsistent in quality and scalability. Therefore, there is a clear need for an evidence-based framework that empirically links generative architecture design with data structure to guide model selection for cybersecurity applications. This is particularly important for cybersecurity research, where data-driven detection systems rely on consistent and realistic traffic generation to ensure the trustworthiness of model evaluation.

\subsection{Lack of an Architectural Selection Framework}
While various methods exist for generating synthetic traffic, ranging from non-AI to AI approaches \cite{figueira2022survey}, a significant scientific limitation remains, i.e., the lack of an evidence-based architectural selection framework. Prior comparative studies have often treated network traffic data as a homogenous entity, focusing on general performance without accounting for the deep differences in underlying data structure. This study argues that the success or failure of a generative model is fundamentally connected to the architectural compatibility between the model and the specific characteristics of the data. For instance, datasets like NSL-KDD \cite{tavallaee2009detailed} are dominated by discrete and categorical features (e.g., flags and protocols), while comparatively modern datasets like CIC-IDS2017 \cite{sharafaldin2018toward} are characterized by continuous, high-dimensional network flow metrics (e.g., bytes per second).
In summary, the challenges of fidelity, utility, and computational complexity can be traced back to a common underlying issue of architectural mismatch. When the generative model’s design does not align with the data’s structure, fidelity deteriorates, utility declines, and scalability fails. This study directly addresses this problem by introducing an architectural selection framework that quantifies and mitigates these mismatches, thereby improving generative performance and computational efficiency.
\subsection{Research Questions}
To address this critical gap, this study systematically evaluates generative models based on their architectural compatibility with varying network data structures. These questions directly address the three core challenges of fidelity, utility, and computational scalability identified earlier. The following research questions guide our investigation:
\begin{enumerate}
    \item Which generative models produce synthetic network traffic data that is both realistic (fidelity) and useful for downstream tasks such as anomaly detection (utility)?
    \item How do dataset characteristics (e.g., categorical/discrete features in NSL KDD vs. continuous-heavy flows in CIC-IDS2017) influence the comparative model performance?
    \item What are the computational trade-offs associated with these methods, and how do these factors govern their suitability for real-world security deployment?
\end{enumerate}
\subsection{Contributions}
This study introduces an Architectural Selection Framework for generating synthetic network traffic, which fundamentally advances model selection beyond traditional benchmarking. Our key contributions explicitly address the pervasive challenges of architectural mismatch and scalability failure, confirming the scientific impact of our empirical findings:
\begin{enumerate}
    \item We propose and validate an evidence-based framework that demonstrates the optimal generative model choice is rigorously dependent on the underlying dataset structure. This contribution provides a systematic, architecturally informed guide for mitigating generative failures in cybersecurity applications.

    \item Our empirical analysis reveals that GAN-based architectures (CTGAN, CopulaGAN) exhibit superior statistical robustness (low variance from $N=20$ runs), consistently achieving the most stable fidelity-utility balance across heterogeneous network data structures. This finding is a validated benchmark for reliable synthetic data generation.

    \item  We provide the rigorous quantification of the high computational barrier for Diffusion Models and Probabilistic Graphical Models on high-volume, tabular network data. This conclusion asserts that the current architectural design of these models is fundamentally impractical for large-scale security simulation and deployment.
    
    \item We present a unified benchmarking of twelve representative synthetic data generation methods, including the implementation of Fidelity Gatekeeper Metrics (DS/Corr). This foundation establishes a reproducible baseline for advancing generative modeling in cybersecurity and explicitly validates the structural integrity of synthetic data.
\end{enumerate}
The source code for this study is publicly available at \url{https://github.com/AdenRajput/Comparative_Analysis.git}.

\subsection{Paper Organization}
The remainder of this paper is structured as follows. Section II reviews existing literature and outlines the architectural gaps in current synthetic data research. Section III details the experimental methodology and generative model implementation. Section IV defines the evaluation metrics used for fidelity and utility analysis. Section V presents empirical results and the composite architectural trade-off analysis. Section VI provides a technical discussion on architecture-guided model selection for IDS deployment. Section VII outlines directions for future work, and Section VIII concludes the paper.
\section{Related Work}
\label{sec2}
The generation of synthetic network traffic data has evolved significantly, moving from rule-based simulations to advanced deep generative models \cite{shi2025comprehensive}. While many studies aim to improve fidelity (the degree to which synthetic data replicates real network traffic statistics and temporal patterns), fewer address the twin issues of utility (downstream task performance, i.e., IDS) and computational scalability (training cost, memory use, deployment feasibility) \cite{stoian2025survey, abdulganiyu2025xidintfl, wang2025towards}. This section reviews prior efforts grouped by architectural class, highlighting how each addresses (or neglects) these three challenges and identifying the gaps that motivate our architectural selection framework.

\subsection{Statistical and Resampling Architectures (Non-AI Methods): Limited Structural Fidelity}
Non‑AI approaches primarily focus on class imbalance (boosting/replecating rare attack classes) and simple probabilistic modelling, offering utility or scalability benefits but frequently falling short on structural fidelity \cite{figueira2022survey, fishi2025comprehensive}. Resampling techniques such as Random Oversampling (ROS) \cite{dina2022effect}, Synthetic Minority Oversampling Technique (SMOTE) \cite{chawla2002smote}, and Adaptive Synthetic Sampling (ADASYN) \cite{he2008adasyn} target class imbalance and are computationally inexpensive; however, they do not model complex feature dependencies or temporal correlations. Probabilistic models (e.g., Gaussian Mixture Models (GMM) \cite{chokwitthaya2020applying} and Bayesian Networks \cite{martins2024generation}) attempt to learn marginal or joint distributions, but typically rely on parametric assumptions (e.g., Gaussianity) and cannot capture high‐dimensional, non-linear traffic dependencies. For example, survey evidence shows non‑AI methods often produce synthetic datasets with correct marginal statistics yet poor fidelity in temporal/flow structure \cite{shi2025comprehensive}. In the cybersecurity traffic domain, this limitation manifests as synthetic datasets that may be balanced but lead to degraded performance in downstream tasks (e.g., intrusion detection) due to weak structural realism \cite{challagundla2025synthetic}. To summarize, these methods offer scalability and improved class balance (utility), but trade off fidelity, and are seldom architecture‐aware relative to heterogeneous data structures \cite{koubeissy2025survey}.

\subsection{Deep Generative Architectures (AI-Based): Pursuing Fidelity and Utility (with Scalability Trade-offs)}
Deep generative models (VAEs, GANs, diffusion) aim to generate data that better mimics the real traffic distribution, thereby improving fidelity, while offering potential utility gains. Nevertheless, they differ in how they handle computational scalability, and relatively few specifically target heterogeneous traffic dataset structures \cite{wang2025towards}.

\subsubsection{Variational Autoencoders (VAEs)}
Encoder–decoder frameworks, such as Tabular VAEs (TVAE), embed mixed-type data into a latent space and reconstruct synthetic samples \cite{abdulganiyu2025xidintfl}. While they can generalize and are often easier to train than GANs, empirical studies indicate VAEs struggle in high‑dimensional network traffic contexts and fail to preserve rare/discrete patterns or temporal dependencies, resulting in reduced fidelity and sometimes compromised utility. For example, synthetic network traffic generated via VAEs may lack subtle joint‐feature correlations critical for anomaly detection \cite{bourou2021review}. However, empirical evaluations of how VAE architectures align with heterogeneous traffic datasets (categorical vs. continuous) remain limited \cite{kiran2023comparative, anshelevich2025synthetic}.

\subsubsection{Generative Adversarial Networks (GANs)}
GAN-based architectures, including Conditional Tabular GAN (CTGAN) \cite{xu2019modeling}, Wasserstein GAN with Gradient Penalty (WGAN-GP) \cite{arjovsky2017wasserstein, gulrajani2017improved}, and Cascaded Tabular GAN (CasTGAN) \cite{alshantti2024castgan}, remain the dominant approaches for generating high-fidelity synthetic tabular data. They are exceptionally prominent in synthetic network traffic research; for example, surveys of GANs for network traffic generation highlight their ability to capture complex joint distributions and mixed feature types \cite{anande2023generative}. In network traffic contexts, specialized GANs for simulating traffic at both flow level and packet level, i.e., FlowGAN \cite{manocchio2021flowgan} and NetShare \cite{yin2022practical}, have demonstrated improved fidelity and downstream utility \cite{yang2024research, bovenzi2025mapping}. These models tend to deliver superior performance for AI model training (anomaly detection, intrusion detection) when synthetic data is used as augmentation. GANs are associated with well-known issues, including training instability, mode collapse, high computational costs, and sensitivity to architecture/hyperparameters. For large network datasets, these costs become non‑trivial. Although many studies compare GAN performance, few analyze why specific GAN architectures succeed on particular traffic dataset types (e.g., categorical vs. continuous). Architectural alignment is rarely discussed \cite{xu2019modeling, rahman2025leveraging}.

\subsubsection{Diffusion Models}
Diffusion models (e.g., TABDDPM \cite{kotelnikov2023tabddpm}, NetDiffus \cite{sivaroopan2024netdiffus}) represent the emerging models in synthetic traffic generation. For example, NetDiffus demonstrates a 66.4\% increase in fidelity and 18.1\% downstream task uplift compared to GAN-based methods \cite{sivaroopan2024netdiffus}. These models can produce highly realistic traffic, including temporal dynamics, through iterative denoising. A key issue is computational scalability,  many time steps, heavy denoising networks, and large memory footprints. Tabular, mixed-type network traffic applications are still in their early stage. A survey of diffusion for tabular data reports promising results but highlights scalability and architecture/data‐mismatch issues \cite{villaizan2025diffusion}. Even fewer studies explicitly consider how diffusion architecture aligns with traffic data structure, especially in cybersecurity contexts \cite{shi2024tabdiff, zhang2023mixed}.

The preceding subsections highlight progress across various generative families; however, these advances remain largely fragmented. Despite improvements in fidelity and utility, a unifying perspective that relates architectural design to dataset structure remains missing.

\subsection{Architectural Compatibility: A Triplet of Fidelity, Utility \& Scalability Gaps}

Despite the escalation of methods, a recurring limitation in the literature is the absence of systematic investigation into how model architecture coheres with the traffic dataset structure (e.g., categorical-heavy vs. continuous-flow-heavy) and how this affects the triad of fidelity, utility, and scalability \cite{figueira2022survey, shi2024tabdiff}. Specifically: 
\begin{itemize}
    \item Few studies analyze how architectural design choices (e.g., conditional generation in GANs vs. iterative denoising in diffusion) affect the fidelity‑utility trade‑off across traffic dataset types \cite{xu2019modeling, sivaroopan2024netdiffus}.
    \item There is a scarcity of standardized benchmarks that concurrently evaluate fidelity, utility, and computational scalability across generative model classes on heterogeneous network traffic datasets \cite{lu2024overlapping}.
    \item Practitioners lack a reliable architectural selection framework that guides model choice based on dataset structure, use case (e.g., intrusion detection), and deployment constraints.
\end{itemize}
In summary, while existing research contributes valuable insights into synthetic traffic generation, none provide a comprehensive, architecture-aware framework that links generative model design to dataset structure and addresses the three core challenges (fidelity, utility, and computational scalability). Our study aims to fill this gap by introducing an evidence‑based architectural selection framework that correlates dataset characteristics with model architecture, empirically evaluates twelve representative methods across heterogeneous traffic datasets, and offers actionable guidance for cybersecurity practitioners.

\section{Experimental Methodology and Generative Model Implementation}
The objective of this methodology is to ensure the comparative analysis adheres to scientific rigor by standardizing the experimental environment and establishing a transparent rationale for model configuration and feature selection. This section describes the properties of the selected datasets, the precise feature selection process, and the specific implementation choices made for each generative model architecture.

\subsection{Dataset Characteristics and Selection Rationale}
To investigate Research Question 2 (Architectural Influence), we selected two prominent network traffic datasets that represent fundamentally distinct data structures and types of attack profiling, i.e., NSL-KDD and CIC-IDS2017.

\begin{table*}[ht]
\centering
\caption{Summary of 26 Encoded selected Features from NSL-KDD Dataset \cite{tavallaee2009detailed}}
\begin{tabular}{|c|l|p{8cm}|}
\hline
\textbf{S.No} & \textbf{Feature} & \textbf{Description} \\ \hline
1  & src\_bytes                 & Number of data bytes from source to destination  \\ \hline
2  & dst\_bytes                 & Number of data bytes from destination to source  \\ \hline
3  & same\_srv\_rate            & Percentage of connections to the same service \\ \hline
4  & diff\_srv\_rate            & Percentage of connections to different services \\ \hline
5  & flag\_SF                   & Connection status with a normal connection (SF: "Normal") \\ \hline
6  & dst\_host\_srv\_count      & Number of connections to the same service as the current connection in the past 100 connections \\ \hline
7  & dst\_host\_same\_srv\_rate & Percentage of connections to the same service for a destination host \\ \hline
8  & logged\_in                 & 1 if successfully logged in; 0 otherwise \\ \hline
9  & dst\_host\_serror\_rate    & Percentage of connections that have ``SYN'' errors \\ \hline
10 & dst\_host\_diff\_srv\_rate & Percentage of connections to different services for a destination host \\ \hline
11 & dst\_host\_srv\_serror\_rate & Percentage of connections that have ``SYN'' errors for a destination host \\ \hline
12 & serror\_rate               & Percentage of connections that have ``SYN'' errors \\ \hline
13 & srv\_serror\_rate          & Percentage of connections that have ``SYN'' errors for the same service \\ \hline
14 & flag\_S0                   & Connection status where no data packets were exchanged (S0: "No Data Exchange") \\ \hline
15 & count                      & Number of connections to the same host as the current connection in the past 2 seconds \\ \hline
16 & service\_http              & HTTP service (1 if used, 0 otherwise) \\ \hline
17 & dst\_host\_srv\_diff\_host\_rate & Percentage of connections to different hosts on the same service \\ \hline
18 & level                      & Threat level of the connection \\ \hline
19 & dst\_host\_count           & Number of connections to the same destination host in the past 100 connections \\ \hline
20 & dst\_host\_same\_src\_port\_rate & Percentage of connections with the same source port to the destination host \\ \hline
21 & service\_private           & Private network service (1 if used, 0 otherwise) \\ \hline
22 & srv\_diff\_host\_rate      & Percentage of connections to different hosts for the same service \\ \hline
23 & srv\_count                 & Number of connections to the same service as the current connection in the past 2 seconds \\ \hline
24 & dst\_host\_srv\_rerror\_rate & Percentage of connections that have ``REJ'' errors for a destination host \\ \hline
25 & service\_domain\_u         & Domain name service (DNS) (1 if used, 0 otherwise) \\ \hline
26 & target                     & class label indicating if the connection is normal or an attack \\ \hline
\end{tabular}
\label{table:nsl_kdd_features}
\end{table*}
\begin{table*}[ht]
\caption{Summary of 21 selected Features from CIC-IDS2017 Dataset \cite{sharafaldin2018toward}}
\centering
\begin{tabular}{|c|l|p{8cm}|}
\hline
\textbf{S.No} & \textbf{Feature} & \textbf{Description} \\ \hline
1  & Average Packet Size        & Average size of the packets in the flow \\ \hline
2  & Packet Length Std          & Standard deviation of the packet lengths in the flow \\ \hline
3  & Packet Length Variance     & Variance of the packet lengths in the flow \\ \hline
4  & Packet Length Mean         & Mean length of the packets in the flow \\ \hline
5  & Total Length of Bwd Packets & Total number of bytes in backward (Bwd) packets \\ \hline
6  & Subflow Bwd Bytes          & Number of bytes in the backward sub-flow \\ \hline
7  & Destination Port           & Port number of the destination host \\ \hline
8  & Avg Bwd Segment Size       & Average size of backward segments \\ \hline
9  & Bwd Packet Length Mean     & Mean length of backward packets \\ \hline
10 & Init\_Win\_bytes\_forward  & Initial window size in bytes for forward direction \\ \hline
11 & Subflow Fwd Bytes          & Number of bytes in the forward sub-flow \\ \hline
12 & Total Length of Fwd Packets & Total number of bytes in forward (Fwd) packets \\ \hline
13 & Max Packet Length          & Maximum length of a packet in the flow \\ \hline
14 & Bwd Packet Length Max      & Maximum length of a backward packet \\ \hline
15 & Init\_Win\_bytes\_backward & Initial window size in bytes for backward direction \\ \hline
16 & Fwd Packet Length Max      & Maximum length of a forward packet \\ \hline
17 & Fwd Packet Length Mean     & Mean length of forward packets \\ \hline
18 & Avg Fwd Segment Size       & Average size of forward segments \\ \hline
19 & Flow IAT Max               & Maximum time interval between packets in the flow \\ \hline
20 & Flow Bytes/s               & Rate of flow in bytes per second \\ \hline
21 & target                     & class label indicating if the flow is benign or malicious \\ \hline
\end{tabular}

\label{table:cic_ids2017_features}
\end{table*}

\subsubsection{NSL-KDD Dataset}
NSL-KDD, a refined version of the original KDD Cup 1999 dataset, is widely used in cybersecurity to evaluate IDS \cite{tavallaee2009detailed}. This dataset was selected as a representative of older network security benchmarks, characterized by a predominance of categorical and discrete features (e.g., flags, protocols, service types), which require specific architectural handling via encoding to maintain dependencies. For this study, the training portion of the NSL-KDD dataset, originally consisting of 125,973 instances and 42 columns (41 features and 1 (binary) target), was used, and following pre-processing, it contained 26 encoded features (Table \ref{table:nsl_kdd_features}).

\subsubsection{CIC-IDS2017 Dataset}
This data is part of the Canadian Institute for Cybersecurity's intrusion detection dataset collection, which is widely employed for evaluating cybersecurity systems \cite{sharafaldin2018toward}. This contemporary dataset was chosen because it focuses heavily on modern, continuous flow-based features (e.g., Average Packet Size, Flow Bytes/s, Packet Length Variance). This structure tests the models' architectural robustness when learning complex, high-dimensional dependencies inherent in modern flow-level network data. CIC-IDS2017 has one week of captured network traffic data in eight comma-separated files (CSV). These files are generated based on the types of attacks. There are 2,830,743 instances of all eight files and 79 columns (78 numerical features and one binary target). The initial eight raw files were concatenated, cleaned (null, infinity, and the duplicate "Fwd Header Length" column removed), resulting in 21 pre-selected features (Table \ref{table:cic_ids2017_features})

\begin{table}[ht]
\caption{Mutual Information Score of NSL-KDD Features}
\centering
\begin{tabular}{|l|c|}
\hline
\textbf{Feature} & \textbf{Mutual Information Score} \\ \hline
src\_bytes                      & 0.566864 \\ \hline
dst\_bytes                      & 0.439281 \\ \hline
same\_srv\_rate                 & 0.369288 \\ \hline
diff\_srv\_rate                 & 0.361895 \\ \hline
flag\_SF                        & 0.341828 \\ \hline
dst\_host\_srv\_count           & 0.335993 \\ \hline
dst\_host\_same\_srv\_rate      & 0.309832 \\ \hline
logged\_in                      & 0.292075 \\ \hline
dst\_host\_serror\_rate         & 0.286589 \\ \hline
dst\_host\_diff\_srv\_rate      & 0.283874 \\ \hline
dst\_host\_srv\_serror\_rate    & 0.281332 \\ \hline
serror\_rate                    & 0.278666 \\ \hline
srv\_serror\_rate               & 0.269186 \\ \hline
flag\_S0                        & 0.263399 \\ \hline
count                           & 0.262733 \\ \hline
service\_http                   & 0.191343 \\ \hline
dst\_host\_srv\_diff\_host\_rate & 0.189822 \\ \hline
level                           & 0.153819 \\ \hline
dst\_host\_count                & 0.144479 \\ \hline
dst\_host\_same\_src\_port\_rate & 0.131316 \\ \hline
service\_private                & 0.118493 \\ \hline
srv\_diff\_host\_rate           & 0.099441 \\ \hline
srv\_count                      & 0.062476 \\ \hline
dst\_host\_srv\_rerror\_rate    & 0.062244 \\ \hline
service\_domain\_u              & 0.048430 \\ \hline
\end{tabular}
\label{table:nsl-kdd-mutual-info}
\end{table}

\begin{table}[ht]
\caption{Mutual Information Score of CIC-IDS2017 Features}
\centering
\begin{tabular}{|l|c|}
\hline
\textbf{Feature} & \textbf{Mutual Information Score} \\ \hline
Average Packet Size        & 0.347112 \\ \hline
Packet Length Std          & 0.342188 \\ \hline
Packet Length Variance     & 0.342019 \\ \hline
Packet Length Mean         & 0.319635 \\ \hline
Total Length of Bwd Packets & 0.296955 \\ \hline
Subflow Bwd Bytes          & 0.296882 \\ \hline
Destination Port           & 0.291245 \\ \hline
Avg Bwd Segment Size       & 0.287676 \\ \hline
Bwd Packet Length Mean     & 0.287483 \\ \hline
Init\_Win\_bytes\_forward  & 0.287174 \\ \hline
Subflow Fwd Bytes          & 0.284528 \\ \hline
Total Length of Fwd Packets & 0.284264 \\ \hline
Max Packet Length          & 0.264060 \\ \hline
Bwd Packet Length Max      & 0.263210 \\ \hline
Init\_Win\_bytes\_backward & 0.250188 \\ \hline
Fwd Packet Length Max      & 0.246537 \\ \hline
Fwd Packet Length Mean     & 0.213223 \\ \hline
Avg Fwd Segment Size       & 0.213065 \\ \hline
Flow IAT Max               & 0.212817 \\ \hline
Flow Bytes/s               & 0.209784 \\ \hline

\end{tabular}
\label{table:CIC-IDS2017-mutual-info}
\end{table}

\subsection{Mutual Information}
To ensure that the synthetic data generation models are trained on the most impactful features, we employed Mutual Information (MI) \cite{shannon1948mathematical} for feature selection, rather than relying on simpler methods such as Pearson correlation \cite{guyon2003introduction} or opaque tree-based techniques \cite{breiman2001random}.
Correlation was the first and most obvious choice, mathematically defined as the Pearson correlation coefficient (equ: \ref{EQ:CORR}). However, correlation only accounts for linear relationships \cite{guyon2003introduction}. In real-world scenarios, non-linear relationships often need to be identified and quantified to select the optimal set of features.
\begin{equation}
\text{Corr}(X, Y) = \frac{\text{Cov}(X, Y)}{\sigma_X \sigma_Y}
\label{EQ:CORR}
\end{equation}

where \(\text{Cov}(X, Y)\) is the covariance between variables \(X\) and \(Y\), and \(\sigma_X\) and \(\sigma_Y\) are the standard deviations of \(X\) and \(Y\), respectively. While correlation is effective for capturing linear relationships, it does not account for the non-linear relationships that often exist in real-world data, limiting its effectiveness for feature selection.
The second option was a tree-based AI technique, known for its robustness and ability to identify essential features rigorously \cite{breiman2001random}. However, tree-based methods suffer from the "black box" problem, offering little interpretability regarding why certain features are selected.

While approaches like Information Gain (IG) have been used for feature selection, as demonstrated by one study \cite{stiawan2020cicids}, Information Gain only measures the relationship between individual features and the target variable, overlooking interdependencies between features \cite{mahto2023novel}. Mathematically, it measures the reduction in entropy when a feature \(X\) is used to predict the target variable \(Y\).

\begin{equation}
IG(Y, X) = H(Y) - H(Y | X)
\label{eq:IG}
\end{equation}

where \(H(Y)\) is the entropy of the target variable \(Y\), and \(H(Y | X)\) is the conditional entropy of \(Y\) given \(X\). Although Information Gain quantifies the relationship between individual features and the target variable, it ignores interdependencies among the features.

In contrast, MI captures the relationship between features and the target and the dependencies among the features themselves, providing a more holistic evaluation \cite{zhang2012inferring}. MI is defined as:
\begin{equation}
MI(X, Y) = H(X) + H(Y) - H(X, Y)
\label{eq:MI}
\end{equation}

where \(H(X)\) and \(H(Y)\) are the entropies of variables \(X\) and \(Y\), and \(H(X, Y)\) is their joint entropy. By capturing non-linear dependencies and interactions among features, MI offers a more comprehensive framework for feature selection.

Compared to other AI-based feature selection methods, such as recursive feature elimination or embedded methods based on decision trees \cite{hua2009performance}, MI has the upper hand by being model-agnostic. Many AI-based techniques are tied to specific algorithms or models, which may introduce biases or limit generalizability across different datasets or classifiers. MI, however, evaluates feature importance based on statistical dependencies independent of any specific model, making it a versatile and unbiased approach \cite{peng2005feature}. This enables MI to offer a more robust selection process, particularly in scenarios where feature relationships are complex and non-linear, a common occurrence in network data.

In brief, for this study, MI was specifically chosen because it is a model-agnostic metric that quantifies both linear and non-linear dependencies between features and the target variable \cite{church1990word}. This is crucial in network traffic data, where complex, non-linear feature interactions define malicious behavior. By selecting the top 25\% of features based on MI weights (Tables  \ref{table:nsl-kdd-mutual-info} \& \ref{table:CIC-IDS2017-mutual-info}), we ensured that the reduced feature set maintained the core information necessary for accurately representing the data, while reducing computational load. This methodological rigor is essential for the Architectural Selection Framework, as it ensures that any observed difference in architectural competence between models (e.g., GANs vs. VAEs) is due to their inherent design, rather than the initial selection of features.
\begin{figure}[htbp]
\includegraphics[width=\columnwidth]{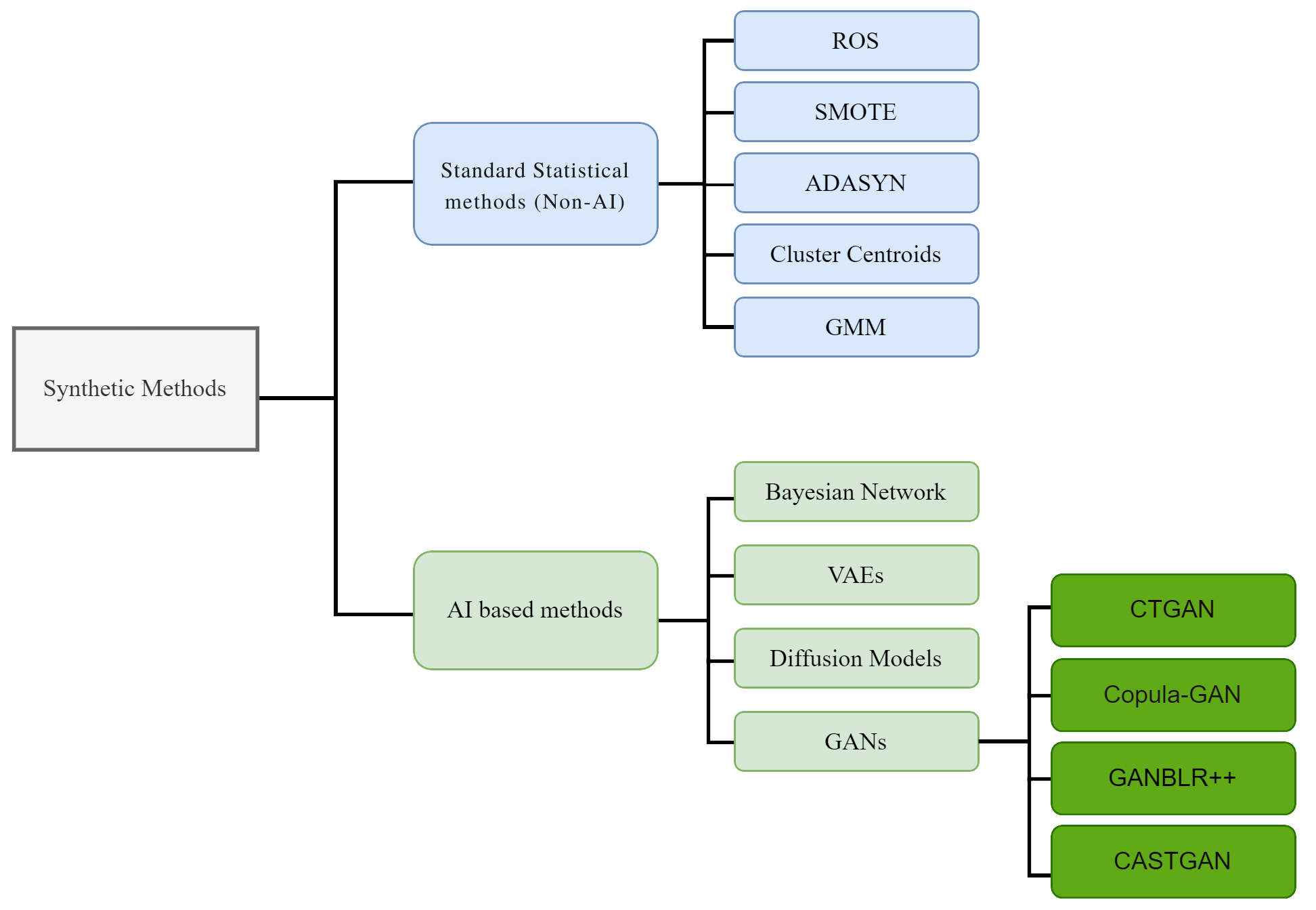}
\caption{Methods for generating synthetic tabular data}
\label{fig:SYN_METHODS}
\end{figure}
\subsection{Generative Model Implementation and Architectural Rationale}
We systematically evaluated twelve methods, grouped into two main architectural categories that split into three (Figure \ref{fig:SYN_METHODS}), to establish a robust comparative framework. Rather than providing general definitions, we focus on the implementation choices that were critical for handling network data complexity. All experiments were conducted on a high-performance workstation utilizing a dedicated NVIDIA GeForce RTX 4090 GPU to ensure uniform computational conditions.
\subsubsection{Standard Statistical and Resampling Architectures (Non-AI Baselines)}
Methods such as ROS, SMOTE, ADASYN, and Cluster Centroids (CC) were included primarily as class balance baselines. Their utility was assessed in correcting class imbalance prior to measuring their ability to capture complex data structures. GMM and Bayesian Networks (BN) were also included to evaluate the performance of classical probabilistic modeling against deep generative methods, especially concerning computational and memory constraints when facing large-scale flow data like CIC-IDS2017.

\subsubsection{AI based methods}
\paragraph{Variational Autoencoders (VAEs) and Diffusion Models}
\begin{itemize}
    \item \textbf{Tabular VAE (TVAE)}: This VAE architecture was implemented to evaluate its ability to learn high-dimensional latent representations of network traffic. We tested whether its inherent encoding-decoding structure, which is computationally expensive, could achieve a superior Fidelity-Utility balance compared to the adversarial process of GANs.
    \item \textbf{Tabular Diffusion Model (TABDDPM)}: This architecture was included to test the viability of a recently emerging paradigm, iterative denoising on structured tabular data. The implementation was specifically challenged against the high volume, unstructured flow data of CIC-IDS2017 to rigorously quantify the architectural limitations related to computational scale.
\end{itemize}
\paragraph{Generative Adversarial Networks (GANs)}
GAN-based models (CTGAN, CopulaGAN, GANBLR++, CasTGAN) were implemented due to their established capability as state-of-the-art synthetic tabular data generators. Their architectures were expected to be robust due to specific design choices. For example, CTGAN and CopulaGAN were implemented utilizing their features designed to handle mixed data types, a necessity for network traffic, which contains both discrete (categorical) and continuous features. CTGAN specifically employs conditional generation and mode-specific normalization, which we hypothesized would be essential for preserving the non-linear feature dependencies crucial for realistic network traffic samples.
\subsection{Hardware and Software Environment}
The experiments were performed on a high-performance workstation (13th Gen Intel® Core™ i9-13900, 32 GB RAM) with a dedicated NVIDIA GeForce RTX 4090 GPU to guarantee reproducibility and provide a fair computational comparison baseline for all AI-based architectures. The software environment utilized Python 3.12.4, leveraging PyTorch for deep learning tasks and the SDV library for tabular GAN implementations (CTGAN, TVAE, CopulaGAN).
To ensure the statistical validity of the performance claims, particularly concerning utility, all TSTR evaluations were executed over twenty independent trials (N=20). This methodology allowed for the calculation of the mean and standard deviation for both Accuracy and F1 score, providing the necessary statistical dispersion required for robust analysis and comparison against the TRTR baseline (as quantified via t-tests in  Table \ref{table:NSL-KDD}, \ref{table:NSLStatisticalSignificance},\ref{table:CIC-IDS2017}, and \ref{table:CICIDS_StatisticalSummary}).

\begin{figure*}[ht]
    \centering
    \includegraphics[width=\textwidth, keepaspectratio]{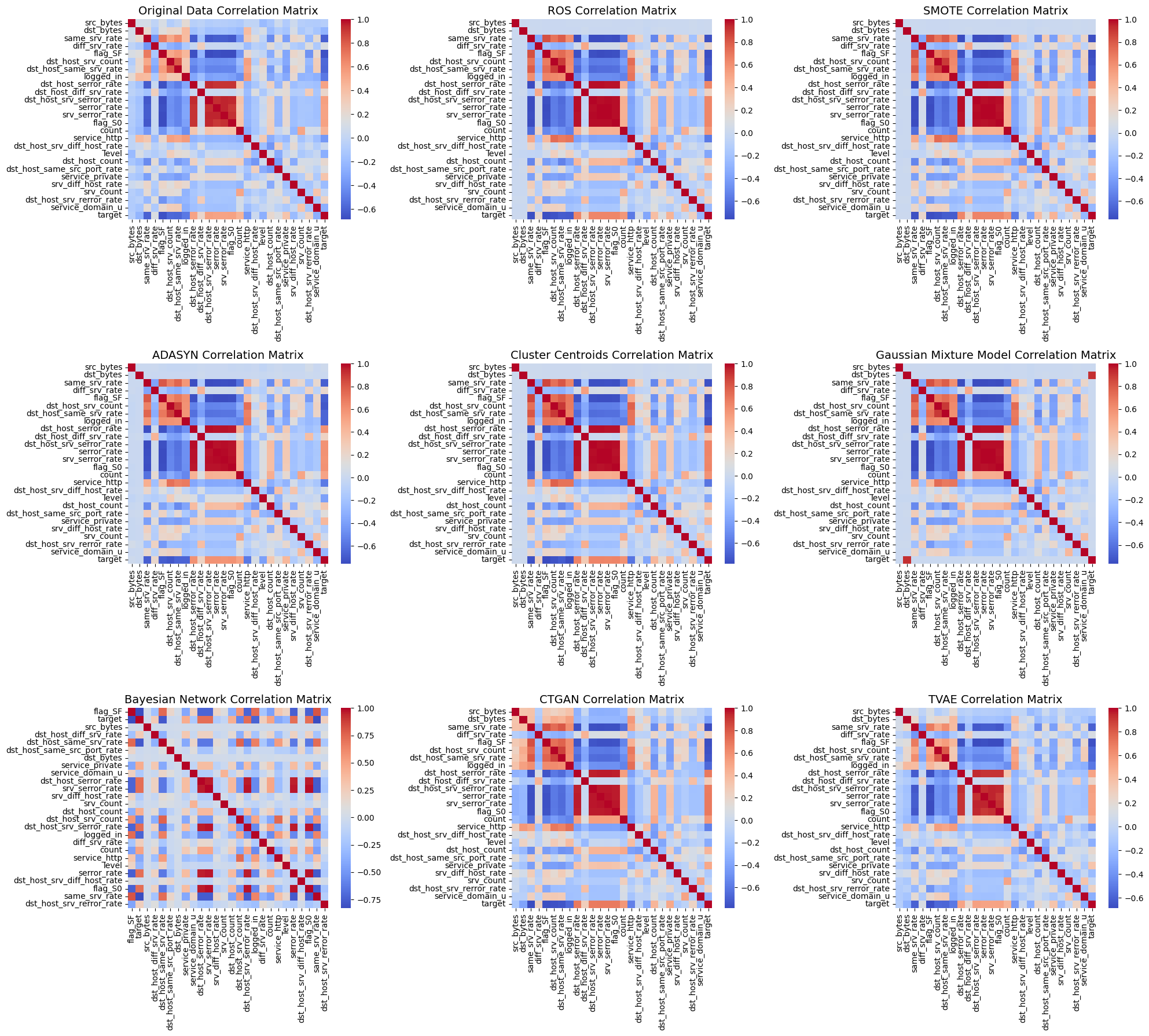}
    \caption{Correlation Heatmap for NSL KDD (Part 1/2)}
    \label{fig:CORR_HEATMAP1}
\end{figure*}

\begin{figure*}[ht]
    \centering
    \includegraphics[width=\textwidth, keepaspectratio]{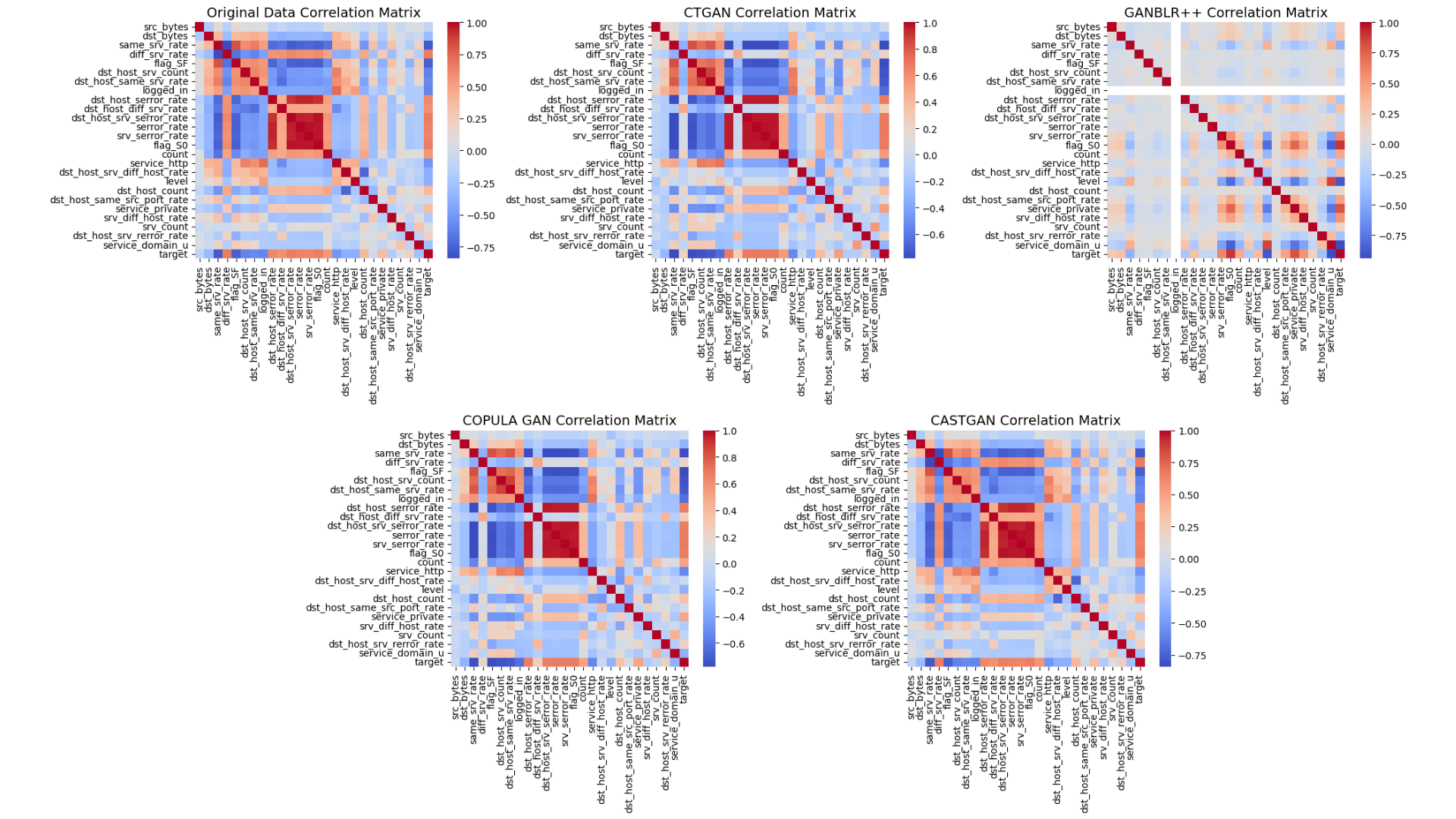}
    \caption{Correlation Heatmap for NSL KDD (Part 2/2)}
    \label{fig:CORR_HEATMAP2}
\end{figure*}

\begin{figure*}[htbp]
    \centering
    \includegraphics[width=\textwidth, height=0.8\textheight, keepaspectratio]{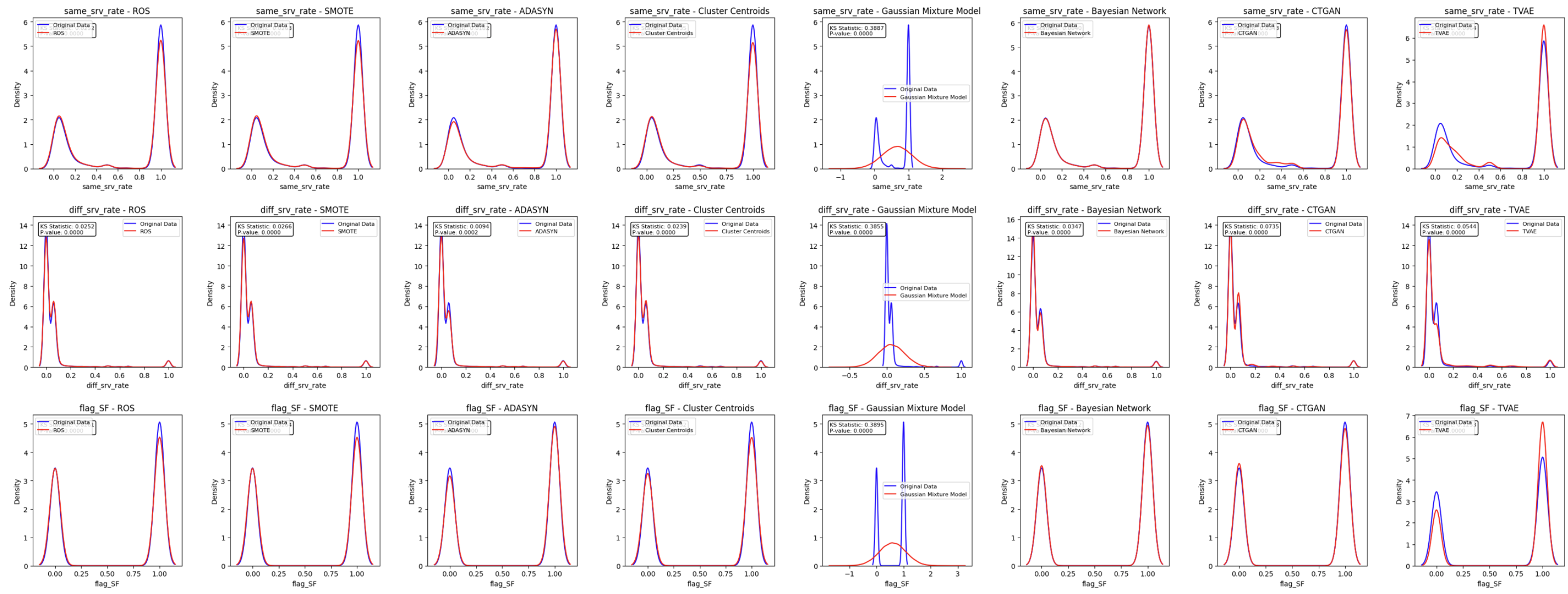}
    \caption{Probability Distribution Comparison for NSL KDD.}
    \label{fig:PD1_NSL_KDD}
\end{figure*}

\section{Evaluation Metrics}
The following section details how each evaluation metric was measured to assess the experimental results.
\subsubsection{Fidelity/Realism Quantification}
Fidelity refers to the necessity for synthetic data to resemble the underlying statistical properties of real data closely \cite{wolf2024benchmarking}. This metric's core mathematical definition is that each variable's probability distribution in the synthetic dataset should closely match that of the corresponding variable in the real dataset \cite{yang2023structured}. However, beyond replicating the individual behavior of each variable, it is equally important to examine the interdependencies and relationships between variables. To evaluate these relationships, we compare the correlations among variables in both datasets \cite{kiran2024methodology}. Thus, we quantified this using three composite metrics:
\begin{itemize}
    \item \textbf{Data Structure (DS)}: This binary metric checks for the strict adherence of synthetic variables to the original data’s logical minimum and maximum values, such as ensuring binary columns (e.g., \texttt{logged\_in}, \texttt{flag\_SF}) contain only 0 or 1 values. Failure to meet this structural requirement is marked as \texttt{NO} and indicates a fundamental flaw in the model's architectural generation process. For example, in  Table \ref{table:NSL-KDD}, the first synthetic data set was generated by ROS, and the value in the DS column is \texttt{YES}, indicating that all binary and boolean variables adhered to their expected values. However, the next row, corresponding to SMOTE, shows \texttt{NO} under the DS column because, apart from the target variable in the NSL-KDD data, the binary and boolean variables did not conform to their expected values. This evaluation method helps assess whether each approach maintains the required data boundaries.
    
    \item \textbf{Correlation (Corr)}: This assessment is performed by examining the correlation heatmap, which represents the absolute values of the correlation coefficients for all variables (see Figure \ref{fig:CORR_HEATMAP1} and Figure \ref{fig:CORR_HEATMAP2} for key NSL KDD heatmaps). It evaluates the strength and direction of relationships in both real and synthetic datasets. If there is a significant difference between the absolute correlation heatmap of the synthetic data and the real data, as well as in the heatmap of the absolute correlation differences, the result is marked as \texttt{NO}. Otherwise, it is marked as \texttt{YES} in the Corr column of all of the results tables.
    
     \item \textbf{Probability Distribution (PD):} This compares the underlying probability distribution of each variable between the real and synthetic datasets, ensuring that their statistical properties are aligned. To calculate the statistics for the PD comparison, the following steps were undertaken:
        \begin{enumerate}
            \item For every variable in the synthetic and real datasets, the underlying PD was estimated using Kernel Density Estimation (KDE).
            \item Each variable's PD was visualized. For clarity, individual figures showing the real and synthetic PDs were generated. In some cases, side-by-side figures were created to provide a direct comparison (Figure \ref{fig:CORR_HEATMAP1}). Complete PD comparison graphs for both NSL-KDD and CIC-IDS2017 are provided in Supplementary material (S1).

            \item The alignment of the PD was quantified using the following formula:
                \begin{equation}
                    \text{PD (\%)} = \frac{\text{Number of Variables with Different PD}}{\text{Total Number of Variables}} \times 100
                     \label{eq:pd_formula}
                \end{equation}
            Where "Number of Variables with Different PD indicates the count of variables whose PD differed between the real and synthetic datasets.
        \end{enumerate}

    For the NSL-KDD dataset, which contains 26 total variables, the results are summarized in Table~\ref{table:NSL-KDD}. Under the PD column, for the ROS method, all variables exhibited identical PDs according to distribution curves in Figure~\ref{fig:PD1_NSL_KDD}, resulting in a PD of $0\%$ difference. Conversely, for the Adaptive Synthetic Sampling method, one variable out of 26 exhibited a different PD (see supplementary material (S1)) for a complete graphical PD comparison. Using Equation~\ref{eq:pd_formula}, the percentage difference was calculated as:
                \begin{equation}
                     \text{PD (\%)} = \frac{1}{26} \times 100 \approx 3.8\%
                \end{equation}

\end{itemize}
\subsubsection{Utility/Classification Performance}
Utility measures the effectiveness of synthetic data for its intended downstream application, training Intrusion IDS models \cite{karr2006framework}. The standard comparison approach compares model performance when trained on real data (TRTR: Train and Test on Real Data) against performance when trained on synthetic data (TSTR: Train on Synthetic Data and Test on Real Data). If performance metrics in the TSTR scenario are comparable to or exceed the TRTR baseline, the synthetic data is considered a viable alternative to the real data \cite{pereira2024assessment}.

For network traffic analysis, utility is determined by assessing the ML model's ability to classify traffic (normal vs. attack). While comprehensive metrics like F1 score, precision, and recall are necessary for imbalanced datasets, the F1 score robustly measures the harmonic mean of precision and recall \cite{zhang2023interpretable}. This study calculated TSTR Accuracy as the primary comparative metric for model utility, as reported in Table \ref{table:NSL-KDD} and Table \ref{table:CIC-IDS2017}. A well-performing TSTR indicates that the synthetic data has successfully captured the essential patterns and relationships within the real data, making it suitable for model development and deployment in privacy-sensitive environments.

To account for the inherent stochasticity of deep generative models and to ensure the scientific rigor of performance differences, all TSTR Accuracy metrics were quantified as the mean ($\bar{x}$) $\pm$ standard deviation ($\sigma$) over twenty independent TSTR runs ($N=20$). Furthermore, we utilized a paired t-test to assess the statistical significance of the performance difference (utility) between key generative models and the TRTR baseline, providing critical, validated evidence for the Architectural Selection Framework.

\subsubsection{Class Balance and Scalability}
Class balance (CB) difference addresses the third research question by measuring the model's ability to maintain a balanced class distribution (Normal vs. Attack) in the generated data. CB refers to the distribution of instances across different classes in the dataset \cite{tholke2023class}. Maintaining a balanced class distribution in real-world datasets is essential for training robust ML models, as imbalanced classes can lead to biased models that perform poorly on underrepresented classes \cite{fernandes2019evolutionary}.
\begin{equation}
    \begin{split}
        \text{Class Balance Difference (\%)} = & \left| P_{\text{synthetic},\text{Normal}} - P_{\text{synthetic},\text{Attack}} \right| \\
        & \times 100
    \end{split}
    \label{eq:class_balance_difference}
\end{equation}
Thus, the CB results mentioned in Table \ref{table:NSL-KDD} and Table \ref{table:CIC-IDS2017} represent how much difference exists between the NORMAL and ATTACK cases. If the classes in the synthetic data are the same, there is zero difference; otherwise, the difference percentage is calculated based on the equation. This equation \ref{eq:class_balance_difference} gives 0\% diff if the normal\% and attack\% classes are equal, i.e., follow a 50-50 ratio in the whole data. Thus, in Table \ref{table:NSL-KDD}, under the CB column for ROS and SMOTE, it is mentioned 0\% diff, and for the Adaptive Synthetic Sampling, it is 0.14\% diff.

For scalability, we assess the computational cost (training time and memory requirements) and failure rates, which are crucial for determining a model's practical suitability for deployment in real-world security infrastructure.
\begin{figure*}[htbp]
    \centering
    \includegraphics[width=\linewidth]{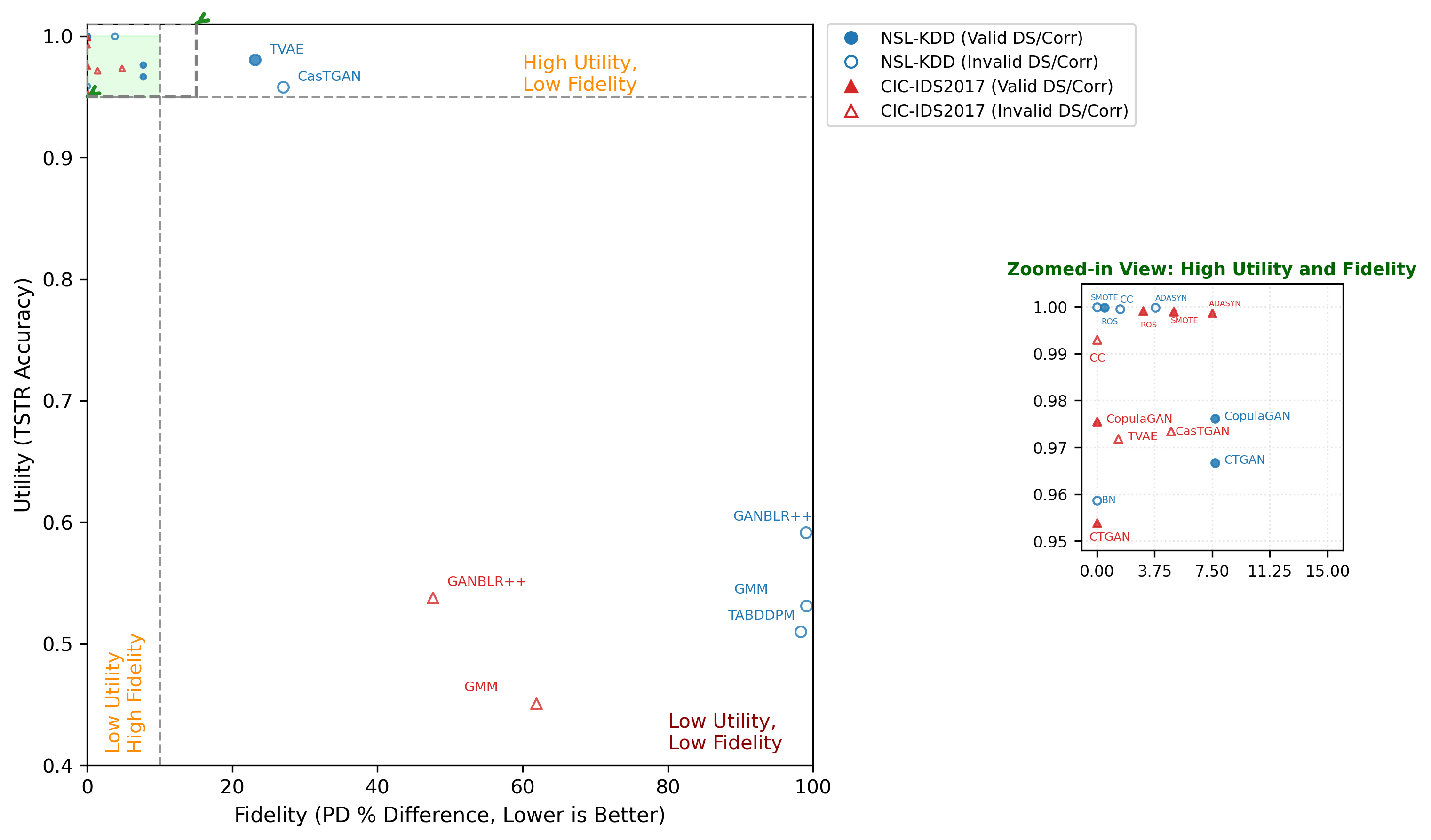}
    \caption{Composite Trade-Off Analysis of Synthetic Data Models}
    \label{fig:composite}
\end{figure*}

\section{Results}
This section presents the empirical results to evaluate the generative model performance based on fidelity, utility, class balance, and scalability across two distinct network traffic data structures (NSL-KDD and CIC-IDS2017). The detailed quantitative results are summarized in Table \ref{table:NSL-KDD} (NSL-KDD) and Table \ref{table:CIC-IDS2017} (CIC-IDS2017), but the primary findings are integrated into a visual architectural trade-off analysis.
\subsection{Architectural Performance on Fidelity and Utility}
We analyze the trade-off between fidelity and utility to answer the first and second research questions, which query the best models and the influence of dataset characteristics. The complete quantitative assessment, including the rigorous F1 score, stability analysis (Mean $\pm$ SD), and statistical significance against the TRTR baseline for NSL-KDD is summarized in Table \ref{table:NSLStatisticalSignificance}, and for the continuous flow-heavy CIC-IDS2017 dataset, the complete statistical summary is provided in Table \ref{table:CICIDS_StatisticalSummary}. A Composite Architectural Trade-Off Plot (Figure \ref{fig:composite}) was generated to visually synthesize the performance of the generative models across both datasets. This figure immediately highlights two key architectural findings:
\begin{enumerate}
    \item \textbf{Dominance of GAN Architectures}: GAN-based models (CTGAN, CopulaGAN) consistently occupy the upper left quadrant across both datasets, demonstrating the most robust balance of high utility and superior fidelity (low PD difference and high correlation preservation) (Table \ref{table:NSL-KDD} and Table \ref{table:CIC-IDS2017}). This architectural superiority is quantified by their performance on CIC-IDS2017, where CopulaGAN achieved an F1 score of $0.9756 \pm 0.0005$, exhibiting extremely low variance in performance and demonstrating its architectural stability when faced with high-volume, continuous-flow data.

    \item \textbf{Dataset Influence on Architectural Failure}: Performance severely degraded for non-GAN architectures when moving from the categorical heavy NSL-KDD to the continuous flow-heavy CIC-IDS2017.
    \begin{itemize}
        \item \textbf{Statistical Failure}: Models like GMM, which rely on parametric assumptions, exhibited the lowest fidelity across both datasets (PD diff of $99.1\%$ on NSL-KDD and $61.9\%$ on CIC-IDS2017), confirming their architectural inability to model the complex, multimodal nature of network flows.
        \item \textbf{Diffusion Model Limitations}: The iterative denoising architecture of TABDDPM performed poorly on NSL-KDD (PD diff $98.3\%$)  and exhibited a catastrophic failure to scale to the CIC-IDS2017 dataset, indicating a fundamental architectural constraint when faced with large-volume, unstructured flow data.
    \end{itemize}
\end{enumerate}
While statistical methods like ROS and SMOTE achieved high Utility (accuracy $0.999$), this is due to their primary function as Class Balance Baselines, generating interpolated samples that, while easily classified, lack the high statistical fidelity required to be considered truly generative. Their inability to maintain data structure (DS: No for SMOTE and ADASYN on NSL-KDD) suggests that their simplicity compromises realism.

\subsection{Computational and Scalability Constraints}
The third research question addresses the practical trade-offs regarding computational cost and scalability for real-world deployment. Our results reveal significant architectural barriers for non-GAN models when scaling to large datasets.
\begin{enumerate}
    \item \textbf{Diffusion Model Failure on CIC-IDS2017}: The most pronounced constraint was observed in the Diffusion Model (TABDDPM) and the Bayesian Network. As shown in Table \ref{table:CIC-IDS2017}, these models failed to execute on the large-scale CIC-IDS2017 dataset due to prohibitively high computational costs and memory requirements for relationship construction. This outcome confirms that architectures relying on extensive iterative processes or graphical model construction are currently impractical for large, high-dimensional tabular network data, regardless of their theoretical generative potential.
    \item \textbf{Probabilistic Model Limitations:} The Bayesian Network (BN) architecture similarly failed to generate results for the CIC-IDS2017 dataset. This highlights the challenge of applying probabilistic graphical models, which require substantial memory and processing power for initial feature relationship construction, to large network flow volumes.
\end{enumerate}
In contrast, CTGAN and CopulaGAN, despite being resource-intensive generative models, successfully processed both datasets, establishing them as the most viable architectural choice when balancing generative power with the need for scalability in a dynamic network security environment.
\subsection{Detailed Comparative Summary}
Tables \ref{table:NSL-KDD} and \ref{table:CIC-IDS2017} provide comprehensive and granular results that support the architectural findings.
\begin{table*}[ht]
\caption{Comparison of Synthetic Data Utility on NSL-KDD Dataset}
\centering
\resizebox{\textwidth}{!}{%
\begin{tabular}{|c|c|c|c|c|c|c|c|}
\hline
\multirow{3}{*}{\textbf{Category}} & \multirow{3}{*}{\textbf{Method}} & \multicolumn{3}{c|}{\textbf{Statistical Similarity (SS)}} & \multirow{3}{*}{\textbf{CB}} & \textbf{Accuracy} & \textbf{F1-Score} \\ \cline{3-5} \cline{7-8}
& & \multirow{2}{*}{\textbf{DS}} & \multirow{2}{*}{\textbf{Corr}} & \multirow{2}{*}{\textbf{PD}} & & \textbf{(Mean $\pm$ SD)} & \textbf{(Mean $\pm$ SD)} \\
& & & & & & (TSTR) & (TSTR) \\ \hline
\multirow{6}{*}{\textbf{Non-AI (Statistical)}}
& \textbf{ROS} & Yes & Yes & 0\% diff & 0\% diff & 0.9998 $\pm$ 0.0001 & 0.9998 $\pm$ 0.0001 \\
\cline{2-8}
& \textbf{SMOTE} & No & Yes & 0\% diff & 0\% diff & 0.9999 $\pm$ 0.0001 & 0.9998 $\pm$ 0.0001 \\
\cline{2-8}
& \textbf{ADASYN} & No & Yes & 3.8\% diff & 0.14\% diff & 0.9998 $\pm$ 0.0001 & 0.9998 $\pm$ 0.0001 \\
\cline{2-8}
& \textbf{Cluster Centroids (CC)} & No & Yes & 0\% diff & 0\% diff & 0.9995 $\pm$ 0.0001 & 0.9998 $\pm$ 0.0001 \\
\cline{2-8}
& \textbf{GMM} & No & Yes & 99.1\% diff & 99.9\% diff & 0.5314 $\pm$ 0.0024 & 0.3718 $\pm$ 0.0028 \\
\cline{2-8}
 \hline
\multirow{7}{*}{\textbf{AI-Based (Classical + Generative)}}
& \textbf{Bayesian Network (BN)} & No & No & 0\% diff & 17.2\% diff & 0.9587 $\pm$ 0.0009 & 0.9582 $\pm$ 0.0009 \\
\cline{2-8}
& \textbf{TVAE} & Yes & Yes & 23.1\% diff & 26.7\% diff & 0.9807 $\pm$ 0.0009 & 0.9813 $\pm$ 0.0009 \\
\cline{2-8}
& \textbf{TABDDPM} & No & No & 98.3\% diff & 34.2\% diff & 0.51 $\pm$ N/A & N/A \\ 
\cline{2-8}
& \textbf{CTGAN} & Yes & Yes & 7.7\% diff & 6.7\% diff & 0.9667 $\pm$ 0.0010 & 0.9685 $\pm$ 0.0010 \\
\cline{2-8}
& \textbf{GANBLR++} & No & No & 99\% diff & 8.8\% diff & 0.5916 $\pm$ 0.0861 & 0.6327 $\pm$ 0.1251 \\
\cline{2-8}
& \textbf{CopulaGAN} & Yes & Yes & 7.7\% diff & 5.4\% diff & 0.9761 $\pm$ 0.0008 & 0.9770 $\pm$ 0.0008 \\
\cline{2-8}
& \textbf{CasTGAN} & No & Yes & 27\% diff & 11.1\% diff & 0.9582 $\pm$ 0.0027 & 0.9588 $\pm$ 0.0027 \\
\hline
\end{tabular}
}
\par\vspace{0.5cm}
\footnotesize{
"\textbf{SS}" = Statistical Similarity,
"\textbf{DS}" = Data Structure,
"\textbf{Corr}" = Correlation,
"\textbf{PD}" = Probability Distribution,
"\textbf{CB}" = Class Balance.
\textbf{All Accuracy and F1-Score results are Mean $\pm$ Standard Deviation across 20 repetitions (TSTR), compared against the TRTR Baseline ($0.9995 \pm 0.0001$).}
}
\label{table:NSL-KDD}
\end{table*}

\begin{table*}[ht]
\caption{\textbf{Statistical Significance Summary of Synthetic Data Utility (TSTR vs. TRTR Baseline, NSL-KDD, N=20)}}
\centering
\begin{tabular}{|l|l|c|c|p{4.5cm}|} 
\hline
\textbf{Category} & \textbf{Method} & \textbf{Accuracy} & \textbf{F1-Score} & \multirow{2}{*}{\textbf{Interpretation}} \\ \cline{3-4}
& & \textbf{(Mean $\pm$ SD)} & \textbf{(Mean $\pm$ SD)} & \\ \hline
\multirow{5}{*}{\text{Non-AI (Statistical)}}
& \textbf{ROS} & 0.9998 $\pm$ 0.0001 & 0.9998 $\pm$ 0.0001 & Near perfect utility and stability. High Utility achieved as a Class Balance Baseline.
 \\ \cline{2-5}
& \textbf{SMOTE} & 0.9999 $\pm$ 0.0001 & 0.9998 $\pm$ 0.0001 & Highest utility, architectural failure (DS: No) confirms compromised fidelity despite high F1. \\ \cline{2-5}
& \textbf{ADASYN} & 0.9998 $\pm$ 0.0001 & 0.9998 $\pm$ 0.0001 & Near perfect utility and stability, architectural failure (DS: No) indicates compromised fidelity.
 \\ \cline{2-5}
& \textbf{CC} & 0.9995 $\pm$ 0.0001 & 0.9998 $\pm$ 0.0001 & Near perfect utility, architectural failure (DS: No) indicates compromised fidelity.
 \\ \cline{2-5}
& \textbf{GMM} & 0.5314 $\pm$ 0.0024 & 0.3718 $\pm$ 0.0028 & Poor utility, low stability, confirms architectural incompatibility with multimodal network data.
 \\ \hline
\multirow{7}{*}{\text{AI-Based (Classical + Generative)}}
& \textbf{BN} & 0.9587 $\pm$ 0.0009 & 0.9582 $\pm$ 0.0009 & Significant utility loss, high stability (low SD) for a classical model.
 \\ \cline{2-5}
& \textbf{TVAE} & 0.9807 $\pm$ 0.0009 & 0.9813 $\pm$ 0.0009 & Highest generative F1, excellent stability, but achieved with higher fidelity loss than GANs.
 \\ \cline{2-5}
& \textbf{TABDDPM} & 0.51 $\pm$ N/A & N/A & N/A \\ \cline{2-5}
& \textbf{CTGAN} & 0.9667 $\pm$ 0.0010 & 0.9685 $\pm$ 0.0010 & High utility with low variance (robustness), optimal balance with high architectural fidelity.
 \\ \cline{2-5}
& \textbf{GANBLR++} & 0.5916 $\pm$ 0.0861 & 0.6327 $\pm$ 0.1251 & Worst overall utility, extreme architectural instability.
 \\ \cline{2-5}
& \textbf{CopulaGAN} & 0.9761 $\pm$ 0.0008 & 0.9770 $\pm$ 0.0008 & Near TVAE utility with highest architectural stability (lowest ± SD among GANs). Optimal performance frontier model.
 \\ \cline{2-5}
& \textbf{CasTGAN} & 0.9582 $\pm$ 0.0027 & 0.9588 $\pm$ 0.0027 & Acceptable utility, slightly higher variance indicates lower architectural stability than CTGAN/CopulaGAN.
 \\ \hline
\end{tabular}
\par\vspace{0.5cm}
\footnotesize{
\textbf{Note:} All TSTR methods showed a \textbf{statistically significant difference} against the \textbf{TRTR Baseline} ($\text{Accuracy}=0.9995 \pm 0.0001, \text{F1}=0.9994 \pm 0.0001$), with $\mathbf{p < 0.0001}$.
The $\text{Mean} \pm \text{SD}$ values quantify the utility and stability across 20 repetitions.
}
\label{table:NSLStatisticalSignificance}
\end{table*}

\begin{table*}[ht]
\caption{Comparison of Synthetic Data Utility on CIC-IDS2017 Dataset}
\centering
\resizebox{\textwidth}{!}{%
\begin{tabular}{|c|c|c|c|c|c|c|c|}
\hline
\multirow{3}{*}{\textbf{Category}} & \multirow{3}{*}{\textbf{Method}} & \multicolumn{3}{c|}{\textbf{Statistical Similarity (SS)}} & \multirow{3}{*}{\textbf{CB}} & \textbf{Accuracy} & \textbf{F1-Score} \\ \cline{3-5} \cline{7-8}
& & \multirow{2}{*}{\textbf{DS}} & \multirow{2}{*}{\textbf{Corr}} & \multirow{2}{*}{\textbf{PD}} & & \textbf{(Mean $\pm$ SD)} & \textbf{(Mean $\pm$ SD)} \\
& & & & & & (TSTR) & (TSTR) \\ \hline

\multirow{5}{*}{\textbf{Non-AI (Statistical)}}
& \textbf{ROS} & Yes & Yes & 0\% diff & 0\% diff & 0.9991 $\pm$ 0.0001 & 0.9991 $\pm$ 0.0001 \\
\cline{2-8}
& \textbf{SMOTE} & Yes & Yes & 0\% diff & 0\% diff & 0.9990 $\pm$ 0.0004 & 0.9989 $\pm$ 0.0001 \\
\cline{2-8}
& \textbf{ADASYN} & Yes & Yes & 0\% diff & 0.2\% diff & 0.9986 $\pm$ 0.0005 & 0.9986 $\pm$ 0.0003 \\
\cline{2-8}
& \textbf{CC} & No & Yes & 0\% diff & 0\% diff & 0.9930 $\pm$ 0.0001 & 0.9930 $\pm$ 0.0005 \\
\cline{2-8}
& \textbf{GMM} & No & No & 61.9\% diff & 46.1\% diff & 0.4509 $\pm$ 0.0023 & 0.4745 $\pm$ 0.0016 \\ \hline
\multirow{7}{*}{\textbf{AI-Based (Classical + Generative)}}
& \textbf{BN} & \multicolumn{6}{|c|}{\textbf{N/A}} \\ 
\cline{2-8}
& \textbf{TVAE} & Yes & No & 1.4\% diff & 69.4\% diff & 0.9718 $\pm$ 0.0007 & 0.9719 $\pm$ 0.0004 \\
\cline{2-8}
& \textbf{TABDDPM} & \multicolumn{6}{|c|}{\textbf{N/A}} \\ 
\cline{2-8}
& \textbf{CTGAN} & Yes & Yes & 0\% diff & 5\% diff & 0.9538 $\pm$ 0.0003 & 0.9607 $\pm$ 0.0007 \\
\cline{2-8}
& \textbf{GANBLR++} & No & No & 47.6\% diff & 64.4\% diff & 0.5378 $\pm$ 0.0665 & 0.5843 $\pm$ 0.0418 \\
\cline{2-8}
& \textbf{CopulaGAN} & Yes & Yes & 0\% diff & 5\% diff & 0.9755 $\pm$ 0.0005 & 0.9756 $\pm$ 0.0005 \\
\cline{2-8}
& \textbf{CasTGAN} & No & No & 4.8\% diff & 58.3\% diff & 0.9734 $\pm$ 0.0005 & 0.9737 $\pm$ 0.0004 \\
\hline
\end{tabular}
}
\par\vspace{0.5cm}
\footnotesize{
"\textbf{SS}" = Statistical Similarity,
"\textbf{DS}" = Data Structure,
"\textbf{Corr}" = Correlation,
"\textbf{PD}" = Probability Distribution,
"\textbf{CB}" = Class Balance.
\textbf{All Accuracy and F1-Score results are Mean $\pm$ Standard Deviation across 20 repetitions (TSTR), compared against the TRTR Baseline (0.9995).}
}
\label{table:CIC-IDS2017}
\end{table*}

\begin{table*}[ht]
\caption{\textbf{Statistical Significance Summary of Synthetic Data Utility (TSTR vs. TRTR Baseline, CIC-IDS2017, N=20)}}
\centering
\begin{tabular}{|l|l|c|c|p{4.5cm}|} 
\hline
\textbf{Category} & \textbf{Method} & \textbf{Accuracy} & \textbf{F1-Score} & \multirow{2}{*}{\textbf{Interpretation}} \\ \cline{3-4}
& & \textbf{(Mean $\pm$ SD)} & \textbf{(Mean $\pm$ SD)} & \\ \hline
\multirow{5}{*}{\text{Non-AI (Statistical)}}
& \textbf{ROS} & 0.9991 $\pm$ 0.0001 & 0.9991 $\pm$ 0.0001 & Near perfect utility and stability. Highly effective Class Balance Baseline. \\ \cline{2-5}
& \textbf{SMOTE} & 0.9990 $\pm$ 0.0004 & 0.9989 $\pm$ 0.0001 & High utility with minimal variance. Achieves utility via interpolation-based balancing. \\ \cline{2-5}
& \textbf{ADASYN} & 0.9986 $\pm$ 0.0005 & 0.9986 $\pm$ 0.0003 & High utility and stability. Highly effective Class Balance Baseline. \\ \cline{2-5}
& \textbf{CC} & 0.9930 $\pm$ 0.0001 & 0.9930 $\pm$ 0.0005 & High utility and stability, but architectural failure (DS: No) confirms Compromise. \\ \cline{2-5}
& \textbf{GMM} & 0.4509 $\pm$ 0.0023 & 0.4745 $\pm$ 0.0016 & Poor utility, stable but ineffective. Low SD confirms stability, but low F1 confirms unsuitability. \\ \hline
\multirow{7}{*}{\text{AI-Based (Classical + Generative)}}
& \textbf{TVAE} & 0.9718 $\pm$ 0.0007 & 0.9719 $\pm$ 0.0004 & High generative utility and stability (low SD). Utility comes at a cost of fidelity. \\ \cline{2-5}
& \textbf{CTGAN} & 0.9538 $\pm$ 0.0003 & 0.9607 $\pm$ 0.0007 & High utility and stability. Robust architectural performance on continuous flows. \\ \cline{2-5}
& \textbf{GANBLR++} & 0.5378 $\pm$ 0.0665 & 0.5843 $\pm$ 0.0418 & Extreme utility loss coupled with catastrophic architectural instability (high SD). \\ \cline{2-5}
& \textbf{CopulaGAN} & 0.9755 $\pm$ 0.0005 & 0.9756 $\pm$ 0.0005 & Optimal Performance Frontier Model. Highest generative utility and superior stability. \\ \cline{2-5}
& \textbf{CASTGAN} & 0.9734 $\pm$ 0.0005 & 0.9737 $\pm$ 0.0004 & High utility and stability. Architectural failure (DS: No) indicates Compromise.  \\ \hline
\end{tabular}
\par\vspace{0.5cm}
\footnotesize{
\textbf{Note:} The TRTR Baseline for this dataset is $\text{Accuracy}=0.9995$, $\text{F1}=0.9995$. All TSTR methods showed a \textbf{statistically significant difference} against the TRTR Baseline ($\mathbf{p < 0.0001}$).
The $\text{Mean} \pm \text{SD}$ values quantify the utility and stability across 20 repetitions.
}
\label{table:CICIDS_StatisticalSummary}
\end{table*}

The discussion and scientific interpretation of these results in the context of IDS deployment will be fully explored in the discussion section.

\section{Discussion}
The systematic comparison across heterogeneous network datasets and model architectures validates the necessity of an Architectural Selection Framework for generating synthetic network traffic. Our findings extend beyond simple benchmarking to establish evidence-based guidelines for selecting generative models, guided by the critical trade-off between architectural robustness and data structure compatibility.

\subsection{Resolution of the Fidelity-Utility Trade-Off}
Research Question 1 aimed to determine which architectures best strike a balance between realism and practical utility. Figure \ref{fig:composite} definitively shows that GANs, specifically CTGAN and CopulaGAN, form the optimal performance frontier, possessing the necessary architectural complexity to succeed across different network data types (Table \ref{table:NSL-KDD} and Table \ref{table:CIC-IDS2017}). The success of these GAN models is technically rooted in their ability to handle mixed discrete and continuous feature types, such as binary flags, categorical protocols, and continuous flow metrics. CTGAN’s use of a Conditional Generator coupled with mode-specific normalization enables it to preserve complex dependencies, where the adversarial process forces the generator to capture the multivariate dependencies that define realistic attack patterns, resulting in low PD Differences (below the $10\%$ threshold) and passing the stringent Correlation check. Furthermore, the resulting high fidelity translates directly to effective machine learning training, with CopulaGAN achieving a TSTR F1 score of $0.9770 \pm 0.0008$, confirming their viability as replacements for real data in training adaptive IDS.

The structural failure observed in high-utility resampling models (SMOTE, ADASYN, CC) is quantified (Table \ref{table:NSLStatisticalSignificance}), represented by Hollow Markers in Figure \ref{fig:composite}. While these models achieve near-perfect F1 utility ($\approx 0.9998 \pm 0.0001$), confirming their efficacy as class balance baselines, their inability to maintain the fundamental Data Structure (DS: No) constitutes an architectural failure. We define this phenomenon as compromised fidelity; high operational utility is achieved via interpolation at the expense of generative architectural competence. This failure is evident despite the minimal standard deviation, confirming their stability in generating structurally flawed data. While statistical methods achieved high F1 utility ($\approx 0.9990$) on CIC-IDS2017 (Table \ref{table:CICIDS_StatisticalSummary}), their high performance is still architecturally compromised. For instance, CC and CasTGAN exhibit high utility but show Compromised Fidelity by failing the Data Structure check (DS: No in Table \ref{table:CIC-IDS2017}), indicating that their utility is achieved via class manipulation rather than generative competence. Notably, for the continuous flow-heavy CIC-IDS2017 dataset, certain high-fidelity architectures (ROS, SMOTE, CTGAN, and CopulaGAN) achieved a 0\% PD difference (Table \ref{table:CIC-IDS2017}). This zero discrepancy, quantified via Kernel Density Estimation (KDE), serves as empirical proof of the perfect approximation of continuous flow distributions, validating the architectural suitability of these models for modern flow-based security data. Furthermore, the GMM architecture, despite showing a low standard deviation ($±0.0016$), yielded poor utility (F1: $0.4745$) on the CIC-IDS2017 dataset. This confirms that the GMM architecture is stable but ineffective, consistently failing to model the complex, non-Gaussian distributions inherent to continuous flow data.

Among generative architectures, TVAE achieved the highest F1 score ($\mathbf{0.9813 \pm 0.0009}$), demonstrating strong utility. However, the Architectural Selection Framework requires an optimal balance, not just peak utility. TVAE's superior F1 score is achieved at a significant Fidelity cost (PD diff $26.7\%$ vs. CopulaGAN’s $5.4\%$). Conversely, CopulaGAN ($\mathbf{0.9770 \pm 0.0008}$ F1) exhibits the lowest standard deviation among all generative models, proving superior architectural robustness and achieving the ideal balance of high utility with rigorous fidelity preservation, cementing its position on the optimal performance frontier. The poor performance of the iterative Diffusion Model (TABDDPM) on NSL-KDD (PD diff $98.3\%$) and its subsequent failure to scale further highlights the fragility of non-GAN architectures. Finally, the performance of GANBLR++ provides critical evidence of architectural instability, as its low mean utility (F1: $0.6327$ on NSL-KDD, F1: $0.5843$ on CIC-IDS2017) combined with extreme performance variance (up to $\pm 0.1251$ SD on NSL-KDD and $\pm 0.0418$ SD on CIC-IDS2017) makes it statistically unreliable and entirely unsuitable for production IDS deployment.

The statistical summary for the CIC-IDS2017 dataset (Table \ref{table:CICIDS_StatisticalSummary}) further validates the Architectural Robustness of conditional GANs against continuous flow data. CopulaGAN ($0.9756 \pm 0.0005$ F1) maintains the highest generative utility and stability, confirming its architectural suitability for high-dimensional, continuous features. Its minimal standard deviation ($\pm 0.0005$) provides empirical proof that the architectural design is highly robust, resisting the high-variance outputs observed in other unstable architectures, such as GANBLR++ (F1: $0.5843 \pm 0.0418$). The extreme variance of GANBLR++ on both datasets (Tables \ref{table:NSLStatisticalSignificance} and \ref{table:CICIDS_StatisticalSummary}) confirms a consistent architectural failure mode of instability, rendering it unsuitable for reliable security deployment.

\subsection{Architectural Compatibility and Data Structure Influence}
Research Question 2 investigated how the underlying data structure of network traffic influences architectural performance. Our analysis demonstrates that data characteristics are a primary selection constraint.
\begin{itemize}
    \item \textbf{Robustness to Categorical vs. Continuous Data}: The tight clustering of GAN model markers (Blue Circles for categorical NSL-KDD and Red Triangles for continuous CIC-IDS2017 in Figure \ref{fig:composite}) in the top left quadrant provides empirical proof of their architectural resilience. This is quantitatively supported by their consistent stability: CopulaGAN maintains an F1-score standard deviation below $\pm 0.0008$ across both datasets (Table \ref{table:NSLStatisticalSignificance} and Table \ref{table:CICIDS_StatisticalSummary}), demonstrating that its architecture successfully accommodates diverse feature structures, from categorical NSL-KDD to continuous CIC-IDS2017, without performance degradation.
    
    \item \textbf{Vulnerability to Architectural Collapse}: The results expose specific architecture-to-data mismatches. Probabilistic Graphical Models (BN) and Diffusion Models (TABDDPM) demonstrated failure when faced with the high-volume, high-dimensional structure of CIC-IDS2017. This is not merely an optimization problem but an architectural constraint, as the resources required for iterative denoising (TABDDPM) or initial relationship mapping (BN) proved prohibitive for continuous flow data, effectively rendering these approaches non-viable for modern network environments.
\end{itemize}
\subsection{Scalability and Practical Constraints for IDS Deployment}
The computational cost findings directly translate into practical deployment constraints for a security analyst designing an Adaptive IDS (Research Question 3).
\begin{itemize}
    \item \textbf{Impracticality of Emerging Architectures}: The failure of TABDDPM and the Bayesian Network to execute on the large-scale CIC-IDS2017 dataset is the most significant practical finding. As documented in Table \ref{table:CIC-IDS2017} and Table \ref{table:CICIDS_StatisticalSummary}, the utility results for these architectures are labeled 'N/A,' explicitly confirming that the iterative denoising process (TABDDPM) and graphical model construction (BN) pose prohibitive computational barriers for high-volume tabular data. Consequently, these architectures, despite their theoretical promise, cannot be realistically considered for training IDSs that rely on frequent synthetic data generation using massive flow datasets.
    \item \textbf{Operational Feasibility}: CTGAN and CopulaGAN offer the most practical deployment architecture. They successfully navigated both small and large datasets while maintaining high fidelity, establishing a clear path for their implementation in real-world security systems requiring balanced, realistic training data for rapid, adaptive model updates. The trade-off is higher computational demand compared to statistical methods, but this cost is justified by the proven preservation of structural and feature dependencies that statistical methods lack.
\end{itemize}

\section{Future Work}
Building upon the empirical evidence of the Architectural Selection Framework, which proved the dominance of GANs and the scalability failure of Diffusion models for tabular network data, future research must transition to architectural optimization and deployment rigor. We identify three scientifically deep directions:
\subsection{Architectures for Enhanced Fidelity and Privacy}
The current trade-off between realism and privacy requires innovative architectural solutions that guarantee data protection without compromising the high fidelity necessary for anomaly detection.

\begin{itemize}
    \item \textbf{Differential Privacy (DP) Integration in Adversarial Training}: Future work must focus on embedding strict Differential Privacy (DP) mechanisms directly into the generator and discriminator components of CTGAN-like architectures. A key challenge is mitigating the known performance degradation (drop in TSTR accuracy) when using DP. Research should explore novel objective functions that incorporate DP-specific regularizers, aiming to minimize the loss landscape divergence caused by noise injection while maximizing the preservation of complex, non-linear feature dependencies essential for network traffic realism.
    \item \textbf{Architectural Deconvolution for Feature Sensitivity}: We propose investigating architectures capable of quantifying feature sensitivity during the generation process. This involves developing a deconvolutional approach within the GAN latent space to isolate features that pose the highest risk of re-identification (e.g., specific IP addresses or source ports) and applying localized, feature-specific perturbation strategies, rather than applying uniform noise to the entire dataset. This move from global to localized privacy measures is critical for maintaining high utility in sensitive network data.
\end{itemize}
\subsection{Dynamic Feature Dependency Modeling and Advanced Evaluation}
The findings on Correlation and Data Structure failure modes (Figure \ref{fig:composite}, hollow markers) reveal the inadequacy of current metrics to capture true logical relationships in network traffic. 
\begin{itemize}
    \item \textbf{Conditional Dependency Quantification}: Future research must move beyond simple correlation to rigorously quantify logical and conditional dependencies within network flows (e.g., ensuring a "logged\_in" flag is only '1' if the protocol is TCP). This requires developing a new suite of evaluation metrics based on Conditional Mutual Information (CMI) or Probabilistic Graphical Models (PGM) that can provide quantitative 'pass/fail' criteria for generative validity, going beyond the descriptive PD Difference.
    \item \textbf{Architectural Adaptation for Data Dynamics}: Given the rapid evolution of attack vectors, future generative architectures must adapt to data dynamic features—features that change rapidly over time (e.g., flow rate, packet per second). This requires exploring hybrid sequential-tabular models that can incorporate time-series modeling (e.g., Transformers or LSTMs) into the conditional generation process of GANs, ensuring that the synthesized data not only represents static properties but also realistic temporal patterns of evolving threats.
\end{itemize}
\subsection{Large Scale Simulation}
The scalability barriers identified for Diffusion Models necessitate technical solutions for real-world security deployment.

\begin{itemize}
    \item \textbf{Optimizing Iterative Architectures for Tabular Data}: A deep scientific investigation is required to optimize the iterative denoising kernel of Diffusion Models (TABDDPM). This research should focus on replacing or modifying the Gaussian diffusion process with a kernel specifically adapted for discrete or mixed-type tabular data, potentially by leveraging hierarchical structures or quantization methods to reduce the enormous computational cost associated with the current application of continuous domain diffusion to tabular spaces.
    \item \textbf{Adaptive IDS Architectural Integration}: Future work should focus on operationalizing the Architectural Selection Framework by building an Adaptive Intrusion IDS where the generative model (e.g., CTGAN) is an integrated component of the ML pipeline. This involves designing a system where the IDS dynamically monitors its performance drift and automatically triggers the generation of new, balanced, and high-fidelity synthetic data for retraining, effectively making the IDS self-healing against concept drift and emerging threats.
\end{itemize}

\section{Conclusion}
This study presented an architectural selection framework for synthetic network traffic generation, systematically evaluating twelve generative models, including Statistical, VAE, GAN, and Diffusion architectures, across heterogeneous network datasets (categorical-heavy NSL-KDD and continuous flow-heavy CIC-IDS2017). The findings demonstrate that model-data structural compatibility is the key determinant of generative performance. Among the evaluated methods, GAN-based models, particularly CTGAN and CopulaGAN, consistently achieved the best balance between fidelity and utility, showing strong architectural adaptability across both dataset types. In contrast, statistical models compromised structural integrity for class balance (low fidelity), while diffusion-based models, despite high fidelity, suffered from prohibitive computational costs, limiting their scalability for large network traffic. The proposed framework offers actionable guidance for practitioners to select suitable generative architectures based on dataset structure and deployment constraints, thereby contributing to more efficient and reliable synthetic traffic generation for cybersecurity applications. Future research should extend this evaluation to temporal and multimodal datasets, incorporate adaptive model–data matching mechanisms, and investigate optimization strategies to enhance the scalability of diffusion-based models.

\section*{Supplementary Material}
\label{sec:supplementary_info}

Supplementary files are provided to support the findings of this study. The complete set of comparison graphs and supporting analysis is available as a separate single PDF document.
\begin{itemize}
    \item S1: Complete Probability Distribution Comparison for NSL-KDD Dataset
    \item S2: Complete Probability Distribution Comparison for CIC-IDS2017 Dataset
    \item S3: Correlation Heatmaps for CIC-IDS2017 Dataset
\end{itemize}

\section*{Acknowledgment}
The work presented here was mainly financed by the European Celtic+ and Swedish Vinnova project (C2022/1-3)" CISSAN – Collective Intelligence Supported by Security Aware Nodes.

\bibliographystyle{ieeetr}
\bibliography{ref}

@article{layeghy2024benchmarking,
  title={Benchmarking the benchmark—Comparing synthetic and real-world Network IDS datasets},
  author={Layeghy, Siamak and Gallagher, Marcus and Portmann, Marius},
  journal={Journal of Information Security and Applications},
  volume={80},
  pages={103689},
  year={2024},
  publisher={Elsevier}
}

@article{figueira2022survey,
  title={Survey on synthetic data generation, evaluation methods and GANs},
  author={Figueira, Alvaro and Vaz, Bruno},
  journal={Mathematics},
  volume={10},
  number={15},
  pages={2733},
  year={2022},
  publisher={MDPI}
}

@article{bourou2021review,
  title={A review of tabular data synthesis using GANs on an IDS dataset},
  author={Bourou, Stavroula and El Saer, Andreas and Velivassaki, Terpsichori-Helen and Voulkidis, Artemis and Zahariadis, Theodore},
  journal={Information},
  volume={12},
  number={09},
  pages={375},
  year={2021},
  publisher={MDPI}
}

@inproceedings{ganji2023towards,
  title={Towards data generation to alleviate privacy concerns for cybersecurity applications},
  author={Ganji, Dhiraj and Chakraborttii, Chandranil},
  booktitle={2023 IEEE 47th Annual Computers, Software, and Applications Conference (COMPSAC)},
  pages={1447--1452},
  year={2023},
  organization={IEEE}
}

@article{xu2019modeling,
  title={Modeling tabular data using conditional gan},
  author={Xu, Lei and Skoularidou, Maria and Cuesta-Infante, Alfredo and Veeramachaneni, Kalyan},
  journal={Advances in neural information processing systems},
  volume={32},
  year={2019}
}

@article{chawla2002smote,
  title={SMOTE: synthetic minority over-sampling technique},
  author={Chawla, Nitesh V and Bowyer, Kevin W and Hall, Lawrence O and Kegelmeyer, W Philip},
  journal={Journal of artificial intelligence research},
  volume={16},
  pages={321--357},
  year={2002}
}

@inproceedings{he2008adasyn,
  title={ADASYN: Adaptive synthetic sampling approach for imbalanced learning},
  author={He, Haibo and Bai, Yang and Garcia, Edwardo A and Li, Shutao},
  booktitle={2008 IEEE international joint conference on neural networks (IEEE world congress on computational intelligence)},
  pages={1322--1328},
  year={2008},
  organization={Ieee}
}

@inproceedings{chokwitthaya2020applying,
  title={Applying the Gaussian mixture model to generate large synthetic data from a small data set},
  author={Chokwitthaya, Chanachok and Zhu, Yimin and Mukhopadhyay, Supratik and Jafari, Amirhosein},
  booktitle={Construction Research Congress 2020},
  pages={1251--1260},
  year={2020},
  organization={American Society of Civil Engineers Reston, VA}
}

@inproceedings{kotelnikov2023tabddpm,
  title={Tabddpm: Modelling tabular data with diffusion models},
  author={Kotelnikov, Akim and Baranchuk, Dmitry and Rubachev, Ivan and Babenko, Artem},
  booktitle={International Conference on Machine Learning},
  pages={17564--17579},
  year={2023},
  organization={PMLR}
}

@inproceedings{arjovsky2017wasserstein,
  title={Wasserstein generative adversarial networks},
  author={Arjovsky, Martin and Chintala, Soumith and Bottou, L{\'e}on},
  booktitle={International conference on machine learning},
  pages={214--223},
  year={2017},
  organization={PMLR}
}

@inproceedings{tavallaee2009detailed,
  title={A detailed analysis of the KDD CUP 99 data set},
  author={Tavallaee, Mahbod and Bagheri, Ebrahim and Lu, Wei and Ghorbani, Ali A},
  booktitle={2009 IEEE symposium on computational intelligence for security and defense applications},
  pages={1--6},
  year={2009},
  organization={Ieee}
}

@article{alshantti2024castgan,
  title={CasTGAN: Cascaded Generative Adversarial Network for Realistic Tabular Data Synthesis},
  author={Alshantti, Abdallah and Varagnolo, Damiano and Rasheed, Adil and Rahmati, Aria and Westad, Frank},
  journal={IEEE Access},
  year={2024},
  publisher={IEEE}
}

@article{zhang2023interpretable,
  title={Interpretable tabular data generation},
  author={Zhang, Yishuo and Zaidi, Nayyar and Zhou, Jiahui and Li, Gang},
  journal={Knowledge and Information Systems},
  volume={65},
  number={7},
  pages={2935--2963},
  year={2023},
  publisher={Springer}
}

@article{sarker2020cybersecurity,
  title={Cybersecurity data science: an overview from machine learning perspective},
  author={Sarker, Iqbal H and Kayes, ASM and Badsha, Shahriar and Alqahtani, Hamed and Watters, Paul and Ng, Alex},
  journal={Journal of Big data},
  volume={7},
  pages={1--29},
  year={2020},
  publisher={Springer}
}

@article{yang2023structured,
  title={Structured Evaluation of Synthetic Tabular Data},
  author={Yang, Scott Cheng-Hsin and Eaves, Baxter and Schmidt, Michael Thomas and Swanson, Kenneth Brian and Shafto, Patrick},
  year={2023}
}

@article{kiran2024methodology,
  title={A methodology and an empirical analysis to determine the most suitable synthetic data generator},
  author={Kiran, A and Kumar, S Saravana},
  journal={IEEE Access},
  year={2024},
  publisher={IEEE}
}

@article{martins2024generation,
  title={Generation and analysis of synthetic data via Bayesian networks: a robust approach for uncertainty quantification via Bayesian paradigm},
  author={Martins, Larissa NA and Gon{\c{c}}alves, Fl{\'a}vio B and Galletti, Thais P},
  journal={arXiv preprint arXiv:2402.17915},
  year={2024}
}

@article{wolf2024benchmarking,
  title={Benchmarking of synthetic network data: Reviewing challenges and approaches},
  author={Wolf, Maximilian and Tritscher, Julian and Landes, Dieter and Hotho, Andreas and Schl{\"o}r, Daniel},
  journal={Computers \& Security},
  pages={103993},
  year={2024},
  publisher={Elsevier}
}

@article{pereira2024assessment,
  title={Assessment of differentially private synthetic data for utility and fairness in end-to-end machine learning pipelines for tabular data},
  author={Pereira, Mayana and Kshirsagar, Meghana and Mukherjee, Sumit and Dodhia, Rahul and Lavista Ferres, Juan and de Sousa, Rafael},
  journal={Plos one},
  volume={19},
  number={2},
  pages={e0297271},
  year={2024},
  publisher={Public Library of Science San Francisco, CA USA}
}

@article{karr2006framework,
  title={A framework for evaluating the utility of data altered to protect confidentiality},
  author={Karr, Alan F and Kohnen, Christine N and Oganian, Anna and Reiter, Jerome P and Sanil, Ashish P},
  journal={The American Statistician},
  volume={60},
  number={3},
  pages={224--232},
  year={2006},
  publisher={Taylor \& Francis}
}

@article{fernandes2019evolutionary,
  title={Evolutionary inversion of class distribution in overlapping areas for multi-class imbalanced learning},
  author={Fernandes, Everlandio RQ and de Carvalho, Andre CPLF},
  journal={Information Sciences},
  volume={494},
  pages={141--154},
  year={2019},
  publisher={Elsevier}
}

@article{tholke2023class,
  title={Class imbalance should not throw you off balance: Choosing the right classifiers and performance metrics for brain decoding with imbalanced data},
  author={Th{\"o}lke, Philipp and Mantilla-Ramos, Yorguin-Jose and Abdelhedi, Hamza and Maschke, Charlotte and Dehgan, Arthur and Harel, Yann and Kemtur, Anirudha and Berrada, Loubna Mekki and Sahraoui, Myriam and Young, Tammy and others},
  journal={NeuroImage},
  volume={277},
  pages={120253},
  year={2023},
  publisher={Elsevier}
}

@article{gulrajani2017improved,
  title={Improved training of wasserstein gans},
  author={Gulrajani, Ishaan and Ahmed, Faruk and Arjovsky, Martin and Dumoulin, Vincent and Courville, Aaron C},
  journal={Advances in neural information processing systems},
  volume={30},
  year={2017}
}

@article{sharafaldin2018toward,
  title={Toward generating a new intrusion detection dataset and intrusion traffic characterization.},
  author={Sharafaldin, Iman and Lashkari, Arash Habibi and Ghorbani, Ali A and others},
  journal={ICISSp},
  volume={1},
  pages={108--116},
  year={2018}
}

@article{shannon1948mathematical,
  title={A mathematical theory of communication},
  author={Shannon, Claude Elwood},
  journal={The Bell system technical journal},
  volume={27},
  number={3},
  pages={379--423},
  year={1948},
  publisher={Nokia Bell Labs}
}

@article{church1990word,
  title={Word association norms, mutual information, and lexicography},
  author={Church, Kenneth and Hanks, Patrick},
  journal={Computational linguistics},
  volume={16},
  number={1},
  pages={22--29},
  year={1990}
}

@article{stiawan2020cicids,
  title={CICIDS-2017 dataset feature analysis with information gain for anomaly detection},
  author={Stiawan, Deris and Idris, Mohd Yazid Bin and Bamhdi, Alwi M and Budiarto, Rahmat and others},
  journal={IEEE Access},
  volume={8},
  pages={132911--132921},
  year={2020},
  publisher={IEEE}
}

@article{hua2009performance,
  title={Performance of feature-selection methods in the classification of high-dimension data},
  author={Hua, Jianping and Tembe, Waibhav D and Dougherty, Edward R},
  journal={Pattern Recognition},
  volume={42},
  number={3},
  pages={409--424},
  year={2009},
  publisher={Elsevier}
}

@article{guyon2003introduction,
  title={An introduction to variable and feature selection},
  author={Guyon, Isabelle and Elisseeff, Andr{\'e}},
  journal={Journal of machine learning research},
  volume={3},
  number={Mar},
  pages={1157--1182},
  year={2003}
}

@article{breiman2001random,
  title={Random forests},
  author={Breiman, Leo},
  journal={Machine learning},
  volume={45},
  pages={5--32},
  year={2001},
  publisher={Springer}
}

@article{mahto2023novel,
  title={A novel and innovative cancer classification framework through a consecutive utilization of hybrid feature selection},
  author={Mahto, Rajul and Ahmed, Saboor Uddin and Rahman, Rizwan ur and Aziz, Rabia Musheer and Roy, Priyanka and Mallik, Saurav and Li, Aimin and Shah, Mohd Asif},
  journal={BMC bioinformatics},
  volume={24},
  number={1},
  pages={479},
  year={2023},
  publisher={Springer}
}

@article{zhang2012inferring,
  title={Inferring gene regulatory networks from gene expression data by path consistency algorithm based on conditional mutual information},
  author={Zhang, Xiujun and Zhao, Xing-Ming and He, Kun and Lu, Le and Cao, Yongwei and Liu, Jingdong and Hao, Jin-Kao and Liu, Zhi-Ping and Chen, Luonan},
  journal={Bioinformatics},
  volume={28},
  number={1},
  pages={98--104},
  year={2012},
  publisher={Oxford University Press}
}

@article{peng2005feature,
  title={Feature selection based on mutual information criteria of max-dependency, max-relevance, and min-redundancy},
  author={Peng, Hanchuan and Long, Fuhui and Ding, Chris},
  journal={IEEE Transactions on pattern analysis and machine intelligence},
  volume={27},
  number={8},
  pages={1226--1238},
  year={2005},
  publisher={IEEE}
}

@inproceedings{kotal2022privetab,
  title={Privetab: Secure and privacy-preserving sharing of tabular data},
  author={Kotal, Anantaa and Piplai, Aritran and Chukkapalli, Sai Sree Laya and Joshi, Anupam},
  booktitle={Proceedings of the 2022 ACM on International Workshop on Security and Privacy Analytics},
  pages={35--45},
  year={2022}
}

@article{rao2023imbalanced,
  title={An imbalanced generative adversarial network-based approach for network intrusion detection in an imbalanced dataset},
  author={Rao, Yamarthi Narasimha and Suresh Babu, Kunda},
  journal={Sensors},
  volume={23},
  number={1},
  pages={550},
  year={2023},
  publisher={MDPI}
}

@inproceedings{lin2022idsgan,
  title={Idsgan: Generative adversarial networks for attack generation against intrusion detection},
  author={Lin, Zilong and Shi, Yong and Xue, Zhi},
  booktitle={Pacific-asia conference on knowledge discovery and data mining},
  pages={79--91},
  year={2022},
  organization={Springer}
}

@inproceedings{sivaroopan2023synig,
  title={SyNIG: synthetic network traffic generation through time series imaging},
  author={Sivaroopan, Nirhoshan and Madarasingha, Chamara and Muramudalige, Shashika and Jourjon, Guillaume and Jayasumana, Anura and Thilakarathna, Kanchana},
  booktitle={2023 IEEE 48th Conference on Local Computer Networks (LCN)},
  pages={1--9},
  year={2023},
  organization={IEEE}
}

@article{lu2024overlapping,
  title={An overlapping minimization-based over-sampling algorithm for binary imbalanced classification},
  author={Lu, Xuan and Ye, Xuan and Cheng, Yingchao},
  journal={Engineering Applications of Artificial Intelligence},
  volume={133},
  pages={108107},
  year={2024},
  publisher={Elsevier}
}

@article{abdulganiyu2025xidintfl,
  title={XIDINTFL-VAE: XGBoost-based intrusion detection of imbalance network traffic via class-wise focal loss variational autoencoder},
  author={Abdulganiyu, Oluwadamilare Harazeem and Tchakoucht, Taha Ait and Saheed, Yakub Kayode and Ahmed, Hilali Alaoui},
  journal={The Journal of Supercomputing},
  volume={81},
  number={1},
  pages={1--38},
  year={2025},
  publisher={Springer}
}

@article{anande2023generative,
  title={Generative adversarial networks for network traffic feature generation},
  author={Anande, Tertsegha J and Al-Saadi, Sami and Leeson, Mark S},
  journal={International Journal of Computers and Applications},
  volume={45},
  number={4},
  pages={297--305},
  year={2023},
  publisher={Taylor \& Francis}
}

@inproceedings{manocchio2021flowgan,
  title={Flowgan-synthetic network flow generation using generative adversarial networks},
  author={Manocchio, Liam Daly and Layeghy, Siamak and Portmann, Marius},
  booktitle={2021 IEEE 24th International Conference on Computational Science and Engineering (CSE)},
  pages={168--176},
  year={2021},
  organization={IEEE}
}

@inproceedings{yin2022practical,
  title={Practical gan-based synthetic ip header trace generation using netshare},
  author={Yin, Yucheng and Lin, Zinan and Jin, Minhao and Fanti, Giulia and Sekar, Vyas},
  booktitle={Proceedings of the ACM SIGCOMM 2022 Conference},
  pages={458--472},
  year={2022}
}

@article{sivaroopan2024netdiffus,
  title={Netdiffus: Network traffic generation by diffusion models through time-series imaging},
  author={Sivaroopan, Nirhoshan and Bandara, Dumindu and Madarasingha, Chamara and Jourjon, Guillaume and Jayasumana, Anura P and Thilakarathna, Kanchana},
  journal={Computer Networks},
  volume={251},
  pages={110616},
  year={2024},
  publisher={Elsevier}
}

@article{guerra2022datasets,
  title={Datasets are not enough: Challenges in labeling network traffic},
  author={Guerra, Jorge Luis and Catania, Carlos and Veas, Eduardo},
  journal={Computers \& Security},
  volume={120},
  pages={102810},
  year={2022},
  publisher={Elsevier}
}

@article{shahraki2022comparative,
  title={A comparative study on online machine learning techniques for network traffic streams analysis},
  author={Shahraki, Amin and Abbasi, Mahmoud and Taherkordi, Amir and Jurcut, Anca Delia},
  journal={Computer Networks},
  volume={207},
  pages={108836},
  year={2022},
  publisher={Elsevier}
}

@inproceedings{kiran2023comparative,
  title={A comparative analysis of gan and vae based synthetic data generators for high dimensional, imbalanced tabular data},
  author={Kiran, A and Kumar, S Saravana},
  booktitle={2023 2nd International Conference for Innovation in Technology (INOCON)},
  pages={1--6},
  year={2023},
  organization={IEEE}
}

@article{dina2022effect,
  title={Effect of balancing data using synthetic data on the performance of machine learning classifiers for intrusion detection in computer networks},
  author={Dina, Ayesha Siddiqua and Siddique, AB and Manivannan, D},
  journal={IEEE Access},
  volume={10},
  pages={96731--96747},
  year={2022},
  publisher={IEEE}
}

@article{shi2024tabdiff,
  title={Tabdiff: a mixed-type diffusion model for tabular data generation},
  author={Shi, Juntong and Xu, Minkai and Hua, Harper and Zhang, Hengrui and Ermon, Stefano and Leskovec, Jure},
  journal={arXiv preprint arXiv:2410.20626},
  year={2024}
}

@article{zhang2023mixed,
  title={Mixed-type tabular data synthesis with score-based diffusion in latent space},
  author={Zhang, Hengrui and Zhang, Jiani and Srinivasan, Balasubramaniam and Shen, Zhengyuan and Qin, Xiao and Faloutsos, Christos and Rangwala, Huzefa and Karypis, George},
  journal={arXiv preprint arXiv:2310.09656},
  year={2023}
}

@article{shi2025comprehensive,
  title={A comprehensive survey of synthetic tabular data generation},
  author={Shi, Ruxue and Wang, Yili and Du, Mengnan and Shen, Xu and Chang, Yi and Wang, Xin},
  journal={arXiv preprint arXiv:2504.16506},
  year={2025}
}

@article{yang2024research,
  title={Research on the Simulation Method of HTTP Traffic Based on GAN},
  author={Yang, Chenglin and Xu, Dongliang and Ma, Xiao},
  journal={Applied Sciences},
  volume={14},
  number={5},
  pages={2121},
  year={2024},
  publisher={MDPI}
}

@article{villaizan2025diffusion,
  title={Diffusion models for tabular data imputation and synthetic data generation},
  author={Villaiz{\'a}n-Vallelado, Mario and Salvatori, Matteo and Segura, Carlos and Arapakis, Ioannis},
  journal={ACM Transactions on Knowledge Discovery from Data},
  volume={19},
  number={6},
  pages={1--32},
  year={2025},
  publisher={ACM New York, NY}
}

@article{stoian2025survey,
  title={A survey on tabular data generation: Utility, alignment, fidelity, privacy, and beyond},
  author={Stoian, Mihaela C{\"A} and Giunchiglia, Eleonora and Lukasiewicz, Thomas},
  journal={arXiv preprint arXiv:2503.05954},
  year={2025}
}

@article{wang2025towards,
  title={Towards data-centric ai: A comprehensive survey of traditional, reinforcement, and generative approaches for tabular data transformation},
  author={Wang, Dongjie and Huang, Yanyong and Ying, Wangyang and Bai, Haoyue and Gong, Nanxu and Wang, Xinyuan and Dong, Sixun and Zhe, Tao and Liu, Kunpeng and Xiao, Meng and others},
  journal={arXiv preprint arXiv:2501.10555},
  year={2025}
}

@article{challagundla2025synthetic,
  title={Synthetic Tabular Data Generation: A Comparative Survey for Modern Techniques},
  author={Challagundla, Raju and Dorodchi, Mohsen and Wang, Pu and Lee, Minwoo},
  journal={arXiv preprint arXiv:2507.11590},
  year={2025}
}

@article{koubeissy2025survey,
  title={Survey on Tabular Data Privacy and Synthetic Data Generation in Industry 4.0},
  author={Koubeissy, Hadi and Amine, Amir and Kamradt, Marc and Makhoul, Abdallah},
  journal={Applied Intelligence},
  volume={55},
  number={13},
  pages={935},
  year={2025},
  publisher={Springer}
}

@article{anshelevich2025synthetic,
  title={Synthetic tabular data generation using a VAE-GAN architecture},
  author={Anshelevich, Dmitry and Katz, Gilad},
  journal={Knowledge-Based Systems},
  pages={113997},
  year={2025},
  publisher={Elsevier}
}

@article{rahman2025leveraging,
  title={Leveraging gans for synthetic data generation to improve intrusion detection systems},
  author={Rahman, Md Abdur and Francia, Guillermo A and Shahriar, Hossain},
  journal={Journal of Future Artificial Intelligence and Technologies},
  volume={1},
  number={4},
  pages={429--439},
  year={2025}
}

@article{bovenzi2025mapping,
  title={Mapping the landscape of generative AI in network monitoring and management},
  author={Bovenzi, Giampaolo and Cerasuolo, Francesco and Ciuonzo, Domenico and Di Monda, Davide and Guarino, Idio and Montieri, Antonio and Persico, Valerio and Pescap{\'e}, Antonio},
  journal={IEEE Transactions on Network and Service Management},
  year={2025},
  publisher={IEEE}
}

\begin{IEEEbiography}[{\includegraphics[width=1in,height=1.25in,clip,keepaspectratio]{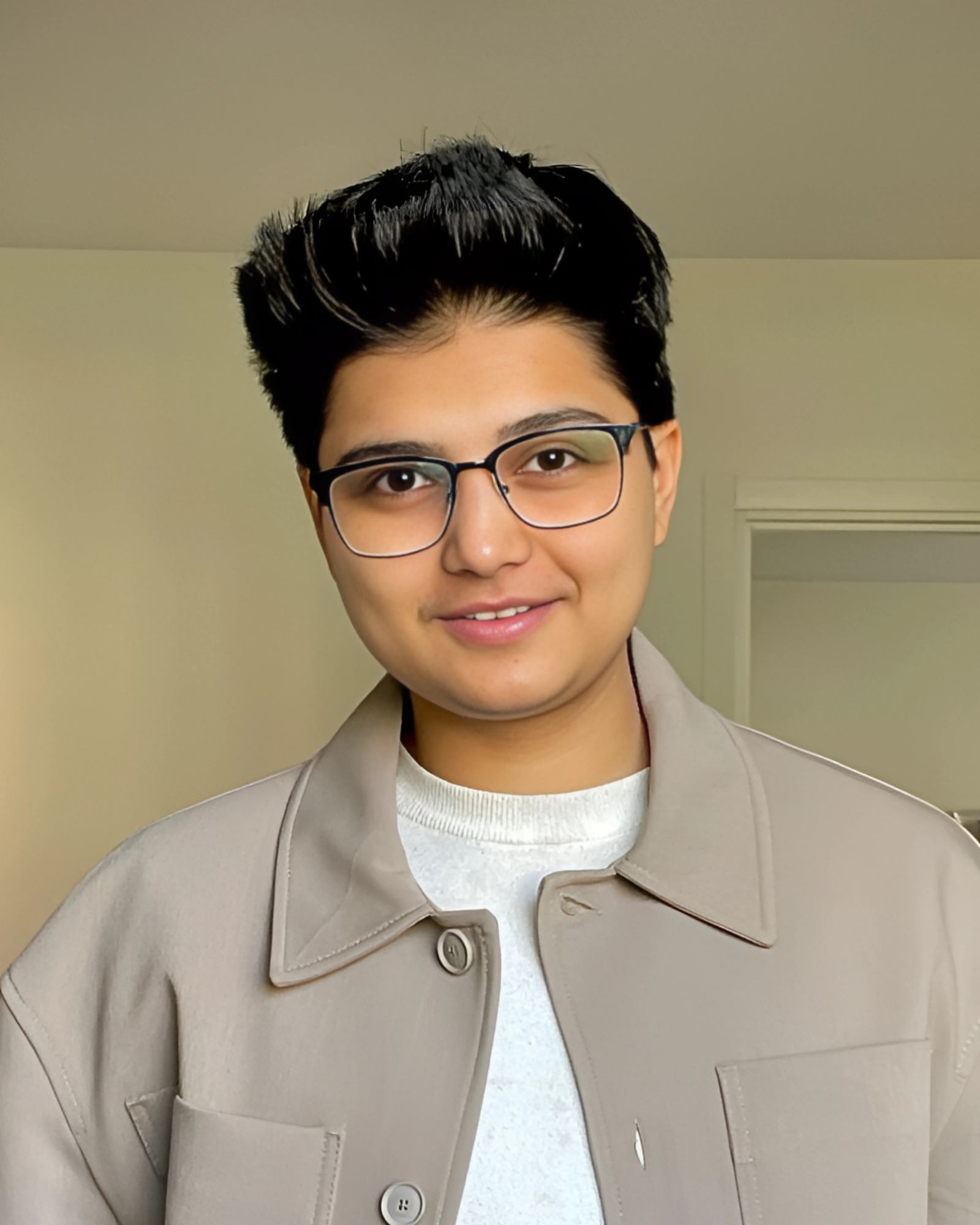}}]{Dure Adan Ammara} is a Ph.D. student in the Department of Computer Science at the Blekinge Institute of Technology (BTH) in Sweden. Adan holds a Bachelor of Science (B.S.) in Mathematics and a Master of Science (M.S.) in Computational Science and Engineering from the National University of Sciences and Technology (NUST) in Pakistan, where he was awarded the President's Gold Medal for academic excellence. His research focuses on applying Generative AI, specifically Generative Adversarial Networks (GANs), to generate synthetic network traffic and smart grid-based Supervisory Control and Data Acquisition (SCADA) data. This research aims to enhance Intrusion Detection Systems within the context of smart grid cybersecurity. It is part of the EU-CISSAN (Celtic Next) project dedicated to advancing cybersecurity solutions for IoT-driven smart grids.
\end{IEEEbiography}

\begin{IEEEbiography}[{\includegraphics[width=1in,height=1.25in,clip,keepaspectratio]{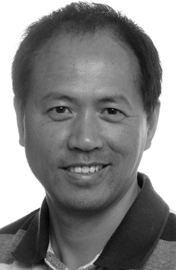}}]{Jianguo Ding} received his Doctorate in Engineering (Dr.-Ing.) from the faculty of mathematics and computer science at the University of Hagen, Germany. He is currently an Associate Professor at the Department of Computer Science, Blekinge Institute of Technology, Sweden. His research interests include cybersecurity, critical infrastructure protection, intelligent technologies, blockchain, distributed systems management and control, and serious games. He is a Senior Member of the IEEE (SM'11) and a Senior Member of the ACM (SM'20).
\end{IEEEbiography}


\begin{IEEEbiography}[{\includegraphics[width=1in,height=1.25in,clip,keepaspectratio]{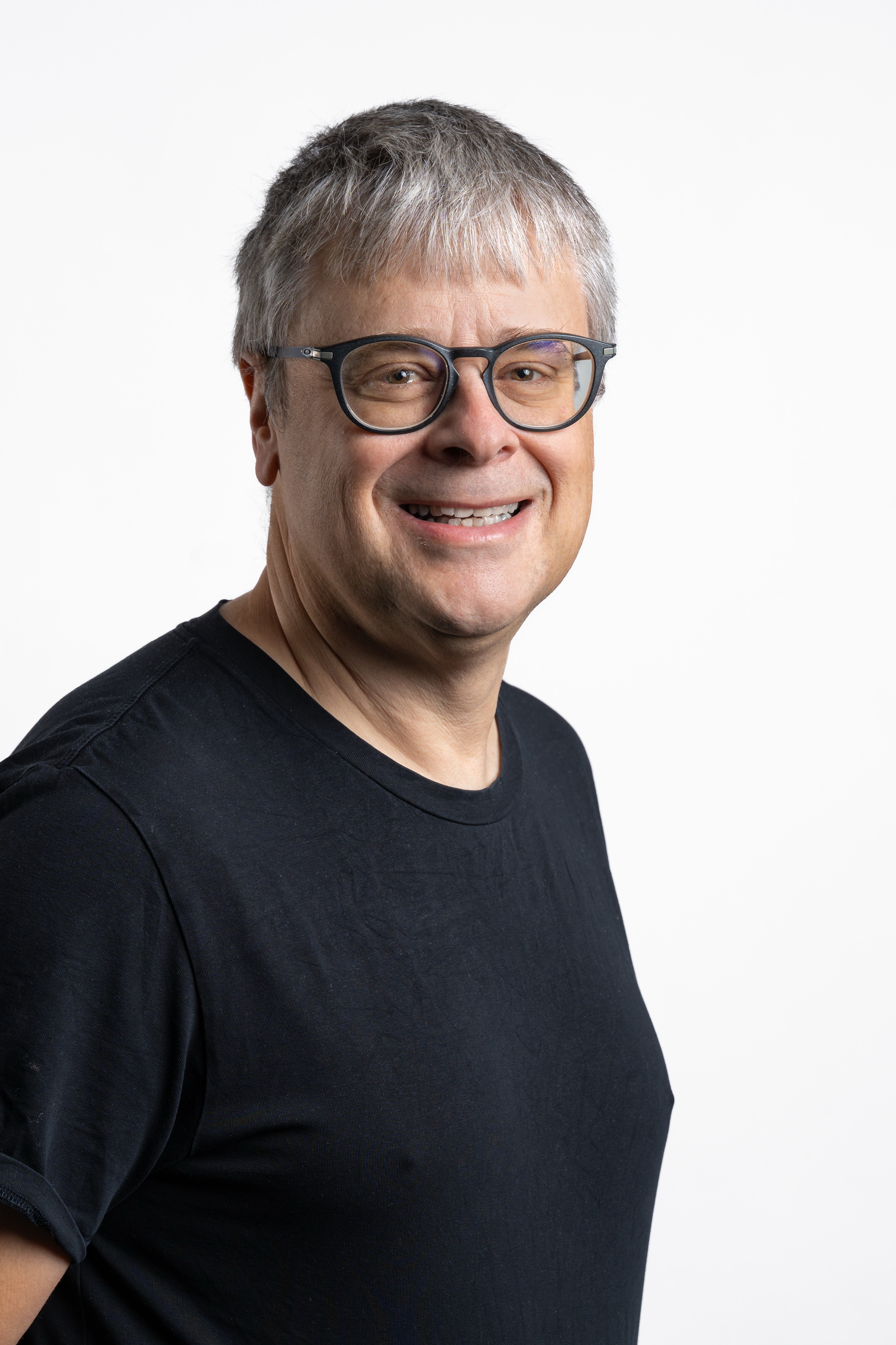}}]{Kurt Tutschku} is a professor for Telecommunication Systems at the Blekinge Institute of Technology (BTH). Prior to BTH, Kurt had the Chair of Future Communication at the Univ. of Vienna (2008 to 2013) and worked at the National Inst. for Inform. and Comm. Tech. (NIST) in Tokyo. He has a Ph.D. in CS ('99) and a Habilitation ('08) from the University of Würzburg, DE.
Kurt's research focuses on efficient and secure architectures and operations of the future. softwarized and smart networks and infrastructures (incl. NFV, SDN, Clouds, SmartGrids, and networked services). He has specialised in their orchestration, performance, and security using technologies, like marketplaces, Blockchains, and distributed AI. Lately, Kurt is addressing the topics of XR, data privacy and digital sovereignty, secure IoT control loops, and the use of generative AI in network security.
He was or is the primary investigator for multiple national and international projects (H2020 Bonseyes/BonsAPPs, HorizonEurope dAIedge, Vinnova/Celtic+ CISSAN, KKS profile HINTS \& KKS HÖG Symphony). He is also the director of studies of the Swedish Industrial Graduate School for Cybersecurity. Kurt served as General Chair of the IEEE NFV-SDN conference from 2017 to 2022.

\end{IEEEbiography}

\EOD
\end{document}